\documentclass[twocolumn,showpacs,preprintnumbers,amsmath,amssymb]{revtex4}

\usepackage{graphicx}
\usepackage{amsmath,amssymb}
\usepackage{dcolumn}
\usepackage{bm}

\begin{document}
\preprint{APS/123-QED}

\title{A proposal of an orbital-dependent correlation energy functional 
for energy-band calculations}

\author{Masahiko Higuchi}
\affiliation{Department of Physics, Faculty of Science, Shinshu University,
Matsumoto 390-8621, Japan}
\author{Hiroshi Yasuhara}%
\affiliation{Department of Physics, Graduate School of Science, Tohoku University,Sendai 980-8578, Japan}

\begin{abstract}
An explicitly orbital-dependent correlation energy functional is proposed,
which is to be used in combination with the orbital-dependent exchange
energy functional in energy-band calculations. It bears a close resemblance
to the second-order direct and exchange perturbation terms calculated with
Kohn-Sham orbitals and Kohn-Sham energies except that one of the two
Coulomb interactions entering each term is replaced by an effective
interaction which contains information about long-, intermediate-, and
short-range correlations beyond second-order perturbation theory.  Such an
effective interaction can rigorously be defined for the
correlation energy of the uniform electron liquid and is evaluated with
high accuracy in order to apply to the orbital-dependent correlation energy
functional. The coupling-constant-averaged spin-parallel and spin-antiparallel
 pair correlation functions are also evaluated
with high accuracy for the electron liquid. The present orbital-dependent
correlation energy functional with the effective interaction borrowed from
the electron liquid is valid for tightly-binding electrons as well as for
nearly-free electrons in marked contrast with the conventional local density
approximation.
\end{abstract}
\maketitle

\section{Introduction}
  The band theory of solids has been developed with the help of the
knowledge of exchange and correlation acquired from the study of the
uniform electron liquid.
The exchange difficulty arising from Hartree-Fock equations has been
overcome by two different methods. One is the standard many-body theory
which has established the quasi-particle picture through Dyson equations
involving the self-energy operator responsible for all exchange and
correlation effects on quasi-particles and clarified the importance
of dynamically screening the exchange potential in the evaluation of the
quasi-particle energy dispersion of the electron liquid \cite{1,2}. The other is the
Slater's intuition \cite{3,4} that the pathological behavior of the
wavenumber-dependence of the Hartree-Fock exchange potential for the
electron liquid around the Fermi wavenumber should be smoothed out by
averaging the potential over all occupied states and that real
many-electron systems may be regarded as a locally uniform electron liquid
with the same local electron density to employ this averaged exchange
potential. The Slater's intuition has developed into a rigorously founded
many-body theory valid for the ground state, namely, density-functional
theory (DFT) \cite{5,6}.

In this paper we propose a new method for further developing the band
theory along DFT. For this purpose we borrow detailed knowledge of long-,
intermediate-, and short-range correlations from the electron liquid, not
such averaged knowledge as the magnitude of its exchange-correlation
energy on which the conventional local density approximation (LDA) is
based. First we give a new expression for the correlation energy of the
electron liquid since it is needed for a proposal of an orbital-dependent
correlation energy functional for use in energy-band calculations. It
resembles second-order direct and exchange perturbation terms, but one of
the two Coulomb interactions entering each term is replaced by an effective
interaction which contains all effects from the higher-order perturbation
terms beyond second-order perturbation theory. The effective interaction
thus defined is reduced in magnitude from the bare Coulomb interaction for
all distances since it takes full account of long-, intermediate-, and
short-range correlations in agreement with the Pauli principle. We evaluate
the effective interaction with high accuracy over the entire region of
metallic densities.

According to Moruzzi, Janak, and Williams \cite{7}, the lowest possible uniform
density  realized in the interstitial region outside muffin-tin spheres of
metals corresponds approximately to the critical density $(r_s=5.25)$ where
the compressibility of the electron liquid becomes divergent, indicative
of an instability of this model. In the construction of an
orbital-dependent correlation energy functional for
energy-band calculations, we think it is physically sound to borrow the
knowledge of the effective interaction from the electron liquid above the
lowest critical density where the system remains thermodynamically stable.
Note that the LDA regards real systems as a locally uniform electron liquid
and borrows its exchange-correlation energy even below the critical
density.

The orbital-dependent correlation energy functional we propose here
consists of  a direct and exchange pair of second-order perturbation like
terms constructed with Kohn-Sham orbitals and Kohn-Sham energies, in which
one of the two Coulomb interactions in each term is replaced by the
effective interaction borrowed from the electron liquid. A combination of
this orbital-dependent correlation energy functional and the
orbital-dependent exchange energy functional is of course reduced to the
LDA in the limit of uniform density. The use of the orbital-dependent
exchange energy functional in DFT makes it possible to cancel exactly the
spurious self-interaction terms involved in the classical Hartree energy
functional. Furthermore, the present orbital-dependent correlation energy
functional cancels exactly higher-order self-interaction terms since it
consists of a direct and exchange pair of terms. We would like to
stress that the present orbital-dependent correlation energy functional
reflects the anti-symmetric property of the many-electron wave function in
a similar way as the Hartree and the exchange energy functionals in the Hartree-Fock
approximation.

In order to obtain the exchange and correlation potential $v_{xc}({\bf r})$ in
DFT, it is necessary to evaluate the functional derivative of the exchange
and correlation energy functional $E_{xc}$ with respect to the electron density
$n({\bf r})$. However, the present correlation energy functional
as well as the exact exchange energy functional is an explicit functional of
Kohn-Sham orbitals, and hence is an implicit functional of $n({\bf r})$. 
The method of evaluating the exchange potential through the functional derivative of the
orbital-dependent exchange energy functional with respect to Kohn-Sham
orbitals has been established and termed the optimized effective
potential method (OEP) \cite{8}.

The present correlation energy functional is given as a functional of
both Kohn-Sham orbitals and the effective interaction borrowed from the
electron liquid. Then, the OEP method can be applied to the evaluation of
the correlation potential but the functional derivative of the effective
interaction with respect to the electron density $n({\bf r})$ has to be treated
separately by resorting to an appropriate approximation. 

Another problem is how to choose the density parameter for the effective potential we borrow
from the electron liquid. In this respect we may utilize the fact that the
compressibility of the electron liquid calculated for the uniform density
in the interstitial region outside muffin-tin spheres is generally a good
approximation to the compressibility of real metals. Hence it is a
reasonable approximation to choose as the density parameter the value appropriate for
the uniform electron density in the interstitial region of metals
calculated from the LDA. The best choice of the density parameter 
is the optimization that minimizes the calculated
ground state energy in the present theory as a function of the density
parameter.

One of the remarkable merits of DFT is that the correlation-induced
reconstruction of Kohn-Sham orbitals accompanied by a change
in the electron density $n({\bf r})$ of the system can in principle be described with the local
correlation potential $v_c({\bf r})$. This can be realized only when accurate information 
about orbital-dependent correlation energy functional $E_c[\{\varphi_i\}]$ and
the potential $v_c({\bf r})$ defined by $\delta E_c[\{\varphi_i\}]/\delta n({\bf r})$ 
is avaiable in addition to the orbital-dependent exchange energy functional $E_x[\{\varphi_i\}]$
and the resulting exact exchange potential $v_x({\bf r})$. 

This reconstruction of Kohn-Sham orbitals is beyond the scope of the LDA since the LDA regards real
systems as a locally uniform electron liquid and borrows its density dependence of
the exchange-correlation energy. The presence of spurious self-interaction terms and the absence of the
correlation-induced reconstruction of Kohn-Sham orbitals are the main two
reasons why the LDA fails to give the correct evaluation of the band gap of
semiconductors and the energy-band structure of the so-called strongly
correlated electron systems.

In the present theory, the exchange and correlation energy functionals are both orbital-dependent and
at the same time they are free from self-interaction errors. The present correlation potential $v_c({\bf r})$ depends
implicitly on the electron density $n({\bf r'})$ at the neighboring position $\bf r'$of $\bf r$ through 
the orbital-dependence of $E_c[\{\varphi_i\}]$.

An important prediction the present theory makes is that
correlation may reduce the screening of the nuclei at short distances
to produce a shrinkage of the electron distribution around individual
nuclei of the system and in accordance with this redistribution of the electron density Kohn-Sham
orbitals may be reconstructed particularly in the neighborhood of the Fermi level or 
the energy gap.

An accurate description of exchange-correlation effects on the
energy-band structure is expected from the present theory. Its
orbital-dependent correlation energy functional, if applied to the
so-called strongly correlated electron systems,  will give rise to 
significant differences in the energy-band structure compared with the case of the
orbital-dependent exchange energy functional alone, particularly in the
neighborhood of the Fermi level or the energy gap.

Since the early stage of the development in DFT a number of authors  \cite{9} have 
been interested in the relationship between
Kohn-Sham equations and Dyson equations. DFT is an exact many-body theory for the ground state
in the framework of the self-consistent one-electron theory. The
Hohenberg-Kohn variational principle leads to a set of Kohn-Sham equations
describing the reference non-interacting system with the same electron
density as the real many-electron system. In this reference system
electrons interact only through the local exchange-correlation potential
$v_{xc}({\bf r} )$. On the other hand, the standard many-body theory is founded on the
variational principle \cite{10,11} that the ground state energy defined as a functional
of the many-body one-electron Green's function $G({\bf r, r'}, \omega)$ 
is stationary with respect to any variation in $G({\bf r, r'}, \omega)$. This stationary
property leads to Dyson equations for the one-electron Green's function. The one-electron Green's
function has the well-founded physical meaning to give the concept of
quasi-particles as low-lying elementary excitations of extended
many-electron systems. 

To make a comparison between Dyson equations and
Kohn-Sham equations we have evaluated with high accuracy the quasi-particle
energy dispersion of the electron liquid over the entire region of metallic
densities. It is found that in the limit of uniform density energy
spectra of Kohn-Sham equations are a rather good approximation to those of
Dyson equations unless they exceed a critical energy beyond which screening no
longer works around the Fermi energy plus the plasmon excitation energy.

In order to justify the application of Kohn-Sham equations to the
band theory it is pointed out that the implicit dependence of the self-energy 
$\Sigma({\bf r,r'},\omega; [G])$ in Dyson equations on the self-consistent Green's function 
$G$ is in one-to-one correspondence to the implicit dependence of exchange-correlation 
potential $v_{xc}({\bf r};[n])$ in Kohn-Sham equations 
on the self-consistent electron density $n({\bf r})$.

The present paper is organized as follows. In Sec. II we give a
detailed analysis of the dielectric formulation for the study of the
electron liquid. This is a section requisite for the introduction of a new
expression for the correlation energy. In Sec. III we calculate the
effective interaction by interpolating between long-range correlation in
the random phase approximation (RPA) and short-range correlation in the
particle-particle ladder approximation such that the exchange counterparts
are self-consistently included. We also calculate the
coupling-constant-averaged spin-parallel and spin-antiparallel pair
correlation functions as well as the correlation energy. In Sec. IV we
propose an orbital-dependent correlation energy functional for use in
energy-band calculations. Some characteristic properties of this
correlation energy functional and how to use it in
practical calculations are discussed. In Sec. V we show an accurate
calculation of the quasi-particle energy dispersion of the electron liquid.
In the last section we give some concluding remarks.
\section{THEORY OF ELECTRON LIQUID}
\subsection{Exchange correction to the RPA}
The random phase approximation (RPA) \cite{12,13,14,15} is a good starting point to describe
long-range correlation characteristic of the electron liquid, but it
violates the Pauli principle since the second- and higher-order exchange
counterparts of the RPA series of perturbation terms are missing. Hubbard \cite{14,15}
was the first who made an attempt to allow for these exchange corrections.
His approximation is an interpolation between long-range correlation in the
RPA and short-range correlation in second-order perturbation theory, but
all the coefficients of third- and higher-order exchange terms in his
approximation are miscounted. The reason for this miscounting is that he
failed to enumerate all the possible proper polarization diagrams leading
to the same third- and higher-order exchange terms. This is not serious for his
evaluation of the ground state energy $E_g$, but is very serious for the purpose of 
evaluating the quasiparticle energy dispersion $E({\bf p})$ and the Landau interaction
function $f^{\sigma \sigma'}({\bf p, p'})$ which are closely related to the first
and the second functional derivations of the ground state energy $E_g[G]$ with
respect to the quasi-particle distribution function $n({\bf p})$, respectively \cite{16}. 

\begin{figure}[tb]
  \includegraphics[width=7.0cm,keepaspectratio]{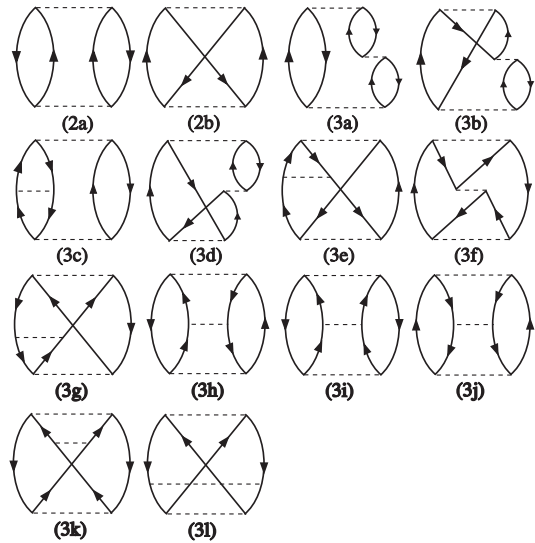}
\caption{\label{fig1} Second- and third-order Goldstone energy diagrams; 
upward and downward solid lines denote particles and holes, respectively,
and dotted lines denote Coulomb interactions. (2a) and (2b): a direct and
exchange pair of second-order diagrams; (3a): third-order PRA diagram; (3b),
(3c) and (3d): singly exchanged diagrams; (3e),(3f) and (3g): doubly
exchanged diagrams; (3h): triply exchanged diagram; (3i) and (3j):
third-order diagrams with a Coulomb interaction line inserted between two
particle or two hole lines in the second-order direct diagram; (3k) and
(3l): exchange counterparts of (3i) and (3j).}
\end{figure}

By way of illustration we consider the exchange counterparts of the
third-order PRA diagram (see Fig. 1). Three different exchange counterparts
can be obtained by exchanging two incoming (or outgoing) particle or hole lines attached
to one of the three Coulomb interaction lines in the RPA diagram.
Additional three different exchange counterparts can also be obtained by
doubly exchanging the PRA diagram. The last exchange counterpart can be
obtained by exchanging all of the three interaction lines; this is a
diagram with one interaction line inserted between a pair of particle and
hole belonging to different bubbles in the second-order direct diagram. The
Hubbard approximation miscounts the coefficients of singly and doubly
exchanged perturbation terms and furthermore neglects the triply exchanged
perturbation term. Besides the eight third-order diagrams we have just
enumerated, there remain more vertex-type diagrams of the same order which
are not derivable from the same order RPA diagram by exchanging, i.e., two
diagrams with one Coulomb interaction line inserted between two particles or two
holes in the second-order direct diagram and their exchange counterparts.
Finally, there are third-order diagrams with a first-order self-energy
correction inserted into the second-order direct and exchange diagrams.

Even if all the possible exchange counterparts of the RPA series are
correctly calculated, another problem arises instead. The resulting
approximation is unable to give the correct description of long-range
correlation characteristic of the electron liquid.

The conserving approximation method by Baym and Kadanoff \cite{17,18} can raise levels of 
approximations maintaining the correct description of
long-range correlation by improving the generating functional $\Phi$ progressively. 
This method guarantees conservation of the local
electron number, the local electron momentum and the local electron energy
as well as the f-sum rule. However, it does not fulfill the Pauli principle 
unless the theory is exact. As a consequence the requirement of the Pauli 
principle in the framework of the conserving approximation method can be a 
guiding principle leading to the exact many-electron theory.
\subsection{Analysis of the dielectric formulation}
\begin{figure}[tb]
  \includegraphics[width=8.5cm,keepaspectratio]{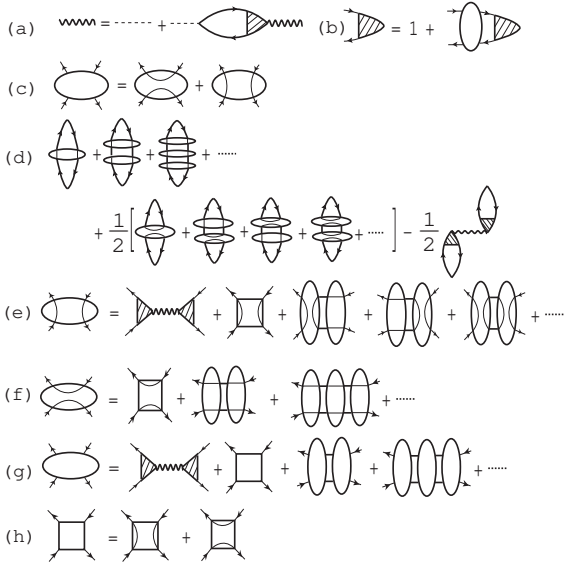}
\caption{\label{fig2} A diagrammatic representation of the integral 
equation for the particle-hole irreducible interaction. (a): the integral equation
for the screened interaction; wavy and dotted lines denote the screened
and the bare Coulomb interactions, respectively, and a bubble with a
cross-hatched triangle denotes the proper polarization function. (b): the
integral equation for the proper vertex part; an ellipse denotes the particle-hole
irreducible interaction. (c): decomposition of the particle-hole irreducible
interaction into two parts, direct and exchange. (d): those polarization
diagrams which integrate to zero due to the Pauli principle; the first infinite
series of diagrams denotes the exchange part of the proper polarization
function excluding the Lindhard function, $\pi_a(q,\omega)-\pi_0(q,\omega)$; the second
infinite series of diagrams with a factor of $1/2$ denotes the direct part of the
proper polarization function, $\pi_b(q,\omega)$ and the third part, two bubbles
connected with a wavy line denotes the improper polarization function. (e)
and (f): the integral equations for the exchange and direct parts of the
particle-hole irreducible interaction, respectively. (g): the integral equation
for the total particle-hole irreducible interaction; the right-hand side
consists of three parts, improper, doubly irreducible, and proper and reducible.
(h): decomposition of the doubly irreducible interaction into two
parts, exchange and direct; a square denotes the doubly irreducible interaction.}
\end{figure}
The dielectric formulation is originally suitable for the description of
long-range correlation of the electron liquid. It discriminates between
proper and improper polarization functions. Improper polarization functions
are responsible for long-range correlation and proper polarization
functions for short-range correlation and exchange. By a diagrammatic
analysis we shall investigate how the requirement of the Pauli principle
should be fulfilled in the dielectric formulation and thereby give the integral
equation for the particle-hole irreducible interaction in a diagrammatic
representation (see Fig. 2).

     We shall start with the following equations \cite{19,20,21,22,23} for the spin-parallel and
spin-antiparallel parts of the static structure factor $S({\bf q})$.
\begin{widetext}
\begin{eqnarray}
S^{\uparrow \uparrow}({\bf q}) &=& \frac{1}{N}\sum_{\sigma}\langle \rho^{\sigma}_{\bf q}\rho^{\sigma}_{\bf -q}\rangle
= \frac{\hbar}{2 \pi n} \int_{0}^{\infty}d \omega {\rm Im}\left\{ \frac{1}{v({\bf q})}
\left(1-\frac{1}{\epsilon({\bf q},\omega)} \right)+\pi_a({\bf q},\omega)\right\} \nonumber \\
&=& \frac{\hbar}{\pi n}\int_{0}^{\infty}d \omega {\rm Im}\left\{\pi_a({\bf q},\omega)+ \frac{1}{2}\pi_b({\bf q},\omega)
-\frac{1}{2}\frac{\pi({\bf q},\omega)v({\bf q})\pi({\bf q},\omega)}{\epsilon({\bf q},\omega)}\right\},\label{eq:2.1}\\
S^{\uparrow \downarrow}({\bf q}) &=& \frac{1}{N}\sum_{\sigma}\langle \rho^{\sigma}_{\bf q}\rho^{-\sigma}_{\bf -q}\rangle
= \frac{\hbar}{2 \pi n} \int_{0}^{\infty}d \omega {\rm Im}\left\{ \frac{1}{v({\bf q})}
\left(1-\frac{1}{\epsilon({\bf q},\omega)} \right)-\pi_a({\bf q},\omega)\right\}\nonumber\\
&=& \frac{\hbar}{\pi n}\int_{0}^{\infty}d \omega {\rm Im}\left\{\frac{1}{2}\pi_b({\bf q},\omega)
-\frac{1}{2}\frac{\pi({\bf q},\omega)v({\bf q})\pi({\bf q},\omega)}
{\epsilon({\bf q},\omega)}\right\},\label{eq:2.2}
\end{eqnarray}
\end{widetext}
where $\langle \cdots \rangle$ denotes the expectation value with respect to the exact 
ground state and $\rho_{\bf q}^{\sigma}$ is the density fluctuation operator; the dielectric
function $\epsilon({\bf q},\omega)$ is defined as 
$\epsilon({\bf q},\omega)=1+v({\bf q})\pi({\bf q},\omega)$ and $v({\bf q})=4\pi e^2/{\bf q}^2$. 
The proper polarization function $\pi({\bf q},\omega)$ is divided into exchange and direct
parts as 
$\pi({\bf q},\omega)=\pi_a({\bf q},\omega)+\pi_b({\bf q},\omega)$ (see Fig. 2 (a), (b) and (c)).
The static structure factor $S({\bf q})$ that gives the description of 
charge-fluctuations is written as
\begin{eqnarray}
S({\bf q})&=&S^{\uparrow \uparrow}({\bf q})+S^{\uparrow \downarrow}({\bf q})\nonumber\\
&=&\frac{1}{N}\langle (\rho^{\uparrow}_{\bf q}+\rho^{\downarrow}_{\bf q})
({\rho^{\uparrow}}_{\bf -q}+{\rho^{\downarrow}}_{\bf -q}) \rangle \nonumber\\
&=&\frac{\hbar}{\pi n}\int_{0}^{\infty}d \omega {\rm Im}\left\{ \frac{1}{v({\bf q})}
\left(1-\frac{1}{\epsilon({\bf q},\omega)}\right)\right\}\label{eq:2.3}.
\end{eqnarray}
 On the other hand, the exchange part of $S({\bf q})$, $\tilde{S}({\bf q})$ 
that gives the description of spin-fluctuations is written as
\begin{eqnarray}
S({\bf q})&=&S^{\uparrow \uparrow}({\bf q})-S^{\uparrow \downarrow}({\bf q})\nonumber\\
&=&\frac{1}{N}\langle (\rho^{\uparrow}_{\bf q}-\rho^{\downarrow}_{\bf q})
(\rho^{\uparrow}_{\bf -q}-\rho^{\downarrow}_{\bf -q}) \rangle \nonumber \\
&=&\frac{\hbar}{\pi n}\int_{0}^{\infty}d \omega~{\rm Im}\{\pi_a({\bf q},\omega)\}\label{eq:2.4}.
\end{eqnarray}
The exchange function $\pi_a({\bf q},\omega)$ comes from those diagrams in which two
external points are connected through particle and hole lines; the function
$\pi_a({\bf q},\omega)$ is spin-dependent and makes a direct contribution to 
$S^{\uparrow \uparrow}({\bf q})$ alone; it is reduced to the Lindhard function 
$\pi_0({\bf q},\omega)$ in the non-interacting limit. The direct function 
$\pi_b({\bf q},\omega)$, on the other hand, comes from those diagrams in which 
two external points are not connected
through particle and hole lines; the function $\pi_b({\bf q},\omega)$ is spin-independent
and makes a contribution to $S^{\uparrow \uparrow}({\bf q})$ and $S^{\uparrow \downarrow}({\bf q})$ equivalently. 
The improper polarization function, $\pi({\bf q},\omega)v({\bf q})\pi({\bf q},\omega)/\epsilon({\bf q},\omega)$, 
whose diagrams can be separated into two pieces by cutting a single Coulomb interaction line
in them, also makes a contribution to $S^{\uparrow \uparrow}({\bf q})$ and $S^{\uparrow \downarrow}({\bf q})$ equivalently.
    
It is important to note that the exchange function $\pi_a({\bf q},\omega)$ makes a
contribution to $S^{\uparrow \uparrow}({\bf q})$ and $S^{\uparrow \downarrow}({\bf q})$ 
indirectly through the improper polarization function common to the two. 
In other words, the two functions $S^{\uparrow \uparrow}({\bf q})$ and $S^{\uparrow \downarrow}({\bf q})$ 
are coupled through $\pi_a({\bf q},\omega)$ in the improper
polarization function. This means that there is correlation between
charge-fluctuations and spin-fluctuations.

The proper polarization function $\pi({\bf q},\omega)$ that is the key quantity in
the dielectric formulation is given as
\begin{eqnarray}
 \pi({\bf q},\omega)&=&-\frac{1}{\Omega}\sum_{{\bf p},\sigma}\int \frac{d \epsilon}{2\pi i}e^{i\delta \epsilon}
G_{\sigma}({\bf{p},\epsilon})G_{\sigma}({\bf{p+q},\epsilon+\omega})\nonumber \\ 
&\times&\Lambda({\bf p},\epsilon;{\bf q},\omega),~~~~~\delta \to +0,\label{eq:2.5}
\end{eqnarray}
where $G_{\sigma}({\bf p},\epsilon)$ is the one-electron Green's function and the proper vertex
part $\Lambda({\bf p},\epsilon;{\bf q},\omega)$ satisfies the following integral equation.
\begin{eqnarray}
 &&\Lambda({\bf p},\epsilon; {\bf q},\omega)\nonumber\\
&=&1+\sum_{\sigma '}\frac{\Omega}{(2\pi)^3}\int d^3{\bf p}'\int\frac{d \epsilon '}{2\pi i}
I^{\sigma \sigma '}({\bf p},\epsilon; {\bf p'},\epsilon'; {\bf q},\omega)\nonumber\\
&&\times G_{\sigma'}({\bf{p}',\epsilon}')G_{\sigma'}({\bf p'+q}, \epsilon '+\omega)\Lambda({\bf p}',\epsilon';{\bf q},\omega).\label{eq:2.6}
\end{eqnarray}
The particle-hole irreducible interaction $I^{\sigma \sigma '}({\bf p}, \epsilon; {\bf p'}, \epsilon';$ ${\bf q},\omega)$ can be
divided into two parts, direct and exchange (see Fig. 2 (c)). The function $\pi_a({\bf q},\omega)$, apart
from $\pi_0({\bf q},\omega)$, is constructed only from exchange parts, whereas the function
$\pi_b({\bf q},\omega)$ is constructed from both direct and exchange parts but must include
at least one direct part.

The spin-parallel and spin-antiparallel pair correlation functions are
given respectively as
\begin{eqnarray}
g^{\uparrow \uparrow}({\bf r})&=&1+\frac{2}{N}\sum_q(S^{\uparrow \uparrow}({\bf q}) -1)e^{i{\bf q \cdot r}},\label{eq:2.7a}\\
g^{\uparrow \downarrow}({\bf r})&=&1+\frac{2}{N}\sum_qS^{\uparrow \downarrow}({\bf q})e^{i{\bf q \cdot r}}.\label{eq:2.7b}
\end{eqnarray}
The Pauli principle means that $g^{\uparrow \uparrow}(0)=0$. This is satisfied only by
$\pi_0({\bf q},\omega)$ or its
modification \cite{19} in which two non-interacting one-electron Green's functions
are replaced by two exact one-electron Green's functions. The Pauli principle
then means that
\begin{eqnarray}
&&\sum_{\bf q}\int_{0}^{\infty} d \omega {\rm Im}\Bigl\{ \pi_a({\bf q},\omega)-\pi_0({\bf q},\omega)
+\frac{1}{2}\pi_b({\bf q},\omega)\nonumber\\
&&~~~~~~~~~~~~~~~-\frac{1}{2}\frac{\pi({\bf q},\omega)v({\bf q})\pi({\bf q},\omega)}{\epsilon({\bf q},\omega)} \Bigl\}=0.\label{eq:2.8}
\end{eqnarray}
We shall see how a direct term and its exchange counterpart which exactly
cancel each other by integrating over ${\bf q}$ and $\omega$ appear in pairs in the
equation above. First of all, the quantity $\pi_a({\bf q},\omega)-\pi_0({\bf q},\omega)$ is nothing but
the exchange counterpart of 
$\frac{1}{2}\{\pi_b({\bf q},\omega)-\pi({\bf q},\omega)v({\bf q})\pi({\bf q},\omega)/\epsilon({\bf q},\omega)\}$
 (see Fig. 2 (d)). 
Consider in detail the interrelation between direct and exchange terms. For this
purpose we shall define the order of $\pi_a({\bf q},\omega)$ and $\pi_b({\bf q},\omega)$ according to the
number of irreducible particle-hole interactions included in them (see Fig. 2 (a), (b) and (c)). It can
easily be seen that the exchange counterpart of the improper term
$-\frac{1}{2}\pi({\bf q},\omega)v({\bf q})\pi({\bf q},\omega)/\epsilon({\bf q},\omega)$ 
is part of the first-order contribution of
$\pi_a({\bf q},\omega)$ that consists of a single exchange part of $I^{\sigma \sigma'}$. The exchange
counterpart of $\frac{1}{2}\pi_b({\bf q},\omega)$ is either the first- or higher-order
contributions of $\pi_a({\bf q},\omega)$. Consider the first-order contribution of
$\frac{1}{2}\pi_b({\bf q},\omega)$ that has a single direct part of $I^{\sigma \sigma'}$. The interaction in it can
be separated into two parts: One is the doubly irreducible interaction and
the other is the direct mate of all the second- and higher-order
contributions of $\pi_a({\bf q},\omega)$ (see Fig. 2 (f)). By doubly irreducible we mean that the diagrams
have no particle-hole pair in the intermediate states when they are seen
not only from bottom to top, but also from left to right. All the second-
and higher-order contributions of $\frac{1}{2}\pi_b({\bf q},\omega)$ include at least one direct
part of $I^{\sigma \sigma'}$. The exchange counterpart of all these contributions is part
of the first-order contribution of $\pi_a({\bf q},\omega)$.

     We have seen how various parts of polarization functions, when
integrated over $\omega$ and $\bf q$, should be cancelled to fulfill the Pauli
principle. From the analysis above we may conclude that the exchange part
of $I^{\sigma \sigma'}$ consists of (i) the exchange counterpart of the interaction
included in the improper polarization function, (ii) the exchange part of
the doubly irreducible interaction, and (iii) the exchange counterpart of
the interaction included in all the second- and higher-order contributions
of $\frac{1}{2}\pi_b({\bf q},\omega)$ (see Fig. 2 (e)). We may also conclude that the direct part of $I^{\sigma \sigma '}$ consists
of (i) the direct part of the doubly irreducible interaction and (ii) the
direct mate of the interaction included in all the second- and higher-order
contributions of $\pi_a({\bf q},\omega)$ (see Fig. 2 (f)). We have thus arrived at the same conclusion
concerning the functional
structure of the particle-hole irreducible interaction that Yasuhara and
Takada \cite{24} obtained from the exact analysis of the self-energy $\Sigma_{\sigma}({\bf p},\epsilon)$ as a
functional of $G_{\sigma '}({\bf p}',\epsilon')$.

     Screening, or long-range correlation in the electron liquid is
ascribed to the long-range nature of the Coulomb interaction between
electrons. An infinite summation of many-body perturbation terms is needed
for the description of screening. As a result, the requirement of the Pauli
principle in the dielectric formulation is fulfilled by the very
complicated form of integral equation for the particle-hole irreducible
interaction diagrammatically shown in Fig. 2 (g). Thus the unified description of screening and exchange is very
much difficult to achieve in the electron liquid.

    The most important part of the doubly irreducible interaction shown in Fig. 2 (h) is 
considered to be an infinite series of particle-particle and
hole-hole ladder interactions (direct and exchange) with respect to the
screened Coulomb interaction. The inclusion of an infinite series of
particle-particle ladder interactions via the bare Coulomb interaction in
the many-body perturbation expansion is indispensable to the proper
description of short-range correlation as well as
to the fulfillment of the cusp condition.
\subsection{Particle-hole irreducible interaction}
The integral equation for 
$I^{\sigma \sigma'}({\bf p},\epsilon;{\bf p}',\epsilon';{\bf q},\omega)$ 
we have just obtained from
a diagrammatic analysis of polarization functions (see Fig. 2 (g)) is quite the
same as the integral equation that Yasuhara and Takada \cite{24} obtained from a
diagrammatic analysis of the self-energy $\Sigma_{\sigma}({\bf p},\epsilon)$ 
as a functional of $G_{\sigma'}({\bf p}',\epsilon')$ in their derivation of 
the symmetric expression for $\Sigma_{\sigma}({\bf p},\epsilon)$ in
which each constituent $G_{\sigma '}$ of $\Sigma_{\sigma}({\bf p},\epsilon)$ 
enters in an equivalent way (see Eq.(\ref{eq:5.2})). According to many-body theory, this coincidence is natural since
the particle-hole irreducible interaction 
$I^{\sigma \sigma'}({\bf p},\epsilon;{\bf p}',\epsilon';{\bf q},\omega)$ in the
limit ${\bf q} = \omega = 0$ is identical to the functional derivative 
$\delta \Sigma_{\sigma}({\bf p},\epsilon)/\delta G_{\sigma'}({\bf p}',\epsilon')$.
Hence the functional structure of $I^{\sigma \sigma'}({\bf p},\epsilon;{\bf p}',\epsilon';{\bf q},\omega)$ 
is identical to that of $\delta \Sigma_{\sigma}({\bf p},\epsilon)/\delta G_{\sigma'}({\bf p}',\epsilon')$.

The conserving approximation method by Baym and Kadanoff \cite{17,18} requires that
an expression for $\Sigma_{\sigma}({\bf p},\epsilon)$ and the one for
$I^{\sigma \sigma'}({\bf p},\epsilon;{\bf p}',\epsilon';{\bf q},\omega)$ 
to be used in the construction of an approximate
proper polarization function $\pi({\bf q},\omega)$ be both $\Phi$-derivable. 
$I^{\sigma \sigma'}({\bf p},\epsilon;{\bf p}',\epsilon';{\bf q},\omega)$ 
has the same functional structure as
$\Sigma_{\sigma}({\bf p},\epsilon)/G_{\sigma'}({\bf p}',\epsilon')$. 
The conserving approximation method then fulfills one of the rigorous
self-consistent requirements imposed on the exact many-body theory.

With the exact generating functional $\Phi$ one can obtain the exact
self-energy and the exact particle-hole irreducible interaction from 
its first and second functional derivatives, respectively.
The standard expression for the self-energy is given as
\begin{eqnarray}
\Sigma_{\sigma}({\bf p},\epsilon)&=&-\frac{1}{\Omega}\sum_{\bf q}\int \frac{d\omega}{2\pi i}
e^{i \delta \omega}G_{\sigma}({\bf p+q},\epsilon+\omega)\nonumber\\
&\times& \frac{v({\bf q})}{\epsilon({\bf q},\omega)}
\Lambda({\bf p},\epsilon; {\bf q},\omega),~~~~\delta\to+0\label{eq:2.9}. 
\end{eqnarray}
 and the one-electron Green's function $G_{\sigma}({\bf p},\epsilon)$ is related to the
self-energy as
\begin{eqnarray}                        
[G_{\sigma}({\bf p},\epsilon)]^{-1}=[G^0_{\sigma}({\bf p},\epsilon)]^{-1}
-\Sigma_{\sigma}({\bf p},\epsilon),\label{eq:2.10}
\end{eqnarray}
where $G^0_{\sigma}({\bf p},\epsilon)$ is the non-interacting one-electron Green's function. 
Note that $\pi({\bf q},\omega)$ and $\Lambda({\bf p},\epsilon; {\bf q},\omega)$ in Eq. (\ref{eq:2.9}) satisfy
Eqs. (\ref{eq:2.5}) and (\ref{eq:2.6}), respectively.  
We would like to stress here the importance of satisfying the integral
equation for the particle-hole irreducible interaction $I^{\sigma \sigma'}({\bf p},\epsilon;{\bf p}',\epsilon';{\bf q},\omega)$ 
in the self-consistent evaluation of various quantities such as
$G_{\sigma}({\bf p},\epsilon)$,
$\Sigma_{\sigma}({\bf p},\epsilon)$, $\Lambda({\bf p},\epsilon; {\bf q},\omega)$, 
$\pi({\bf q},\omega)$ and $\epsilon({\bf q},\omega)$. The accurate knowledge of
$I^{\sigma \sigma'}({\bf p},\epsilon;{\bf p}',\epsilon';{\bf q},\omega)$ makes it possible 
to evaluate all these quantities
self-consistently in agreement with the Pauli principle. In this sense the
integral equation for the particle-hole irreducible interaction is the crux
of many-body theory.

     We have thus clarified that the requirement of the Pauli principle in
the framework of the dielectric formulation can lead to the exact integral
equation for the particle-hole irreducible interaction $I^{\sigma \sigma'}({\bf p},\epsilon;{\bf p}',\epsilon';{\bf q},\omega)$ 
and that the same requirement imposed on the conserving approximation
method can also lead to the exact many-electron theory.

     As can be seen in standard text books \cite{25,26,27,28} on many-body theory, there are
two different expressions for the ground state energy: One is written in
terms of the one-particle Green's function $G_{\sigma}({\bf p},\epsilon)$ and the self-energy
$\Sigma_{\sigma}({\bf p},\epsilon)$ and the other in terms of the dynamically screened Coulomb
interaction $v({\bf q})/\epsilon({\bf q},\omega)$ and the proper polarization function $\pi({\bf q},\omega)$; the
former can easily be transformed into the latter by definition of these
quantities. The two expressions, if inspected into the innermost structure,
can be represented with the same integral equation for the particle-hole
irreducible interaction $I^{\sigma \sigma'}({\bf p},\epsilon;{\bf p}',\epsilon';{\bf q},\omega)$. 
According to the symmetric expression for the self-energy $\Sigma_{\sigma}({\bf p},\epsilon)$ 
by Yasuhara and Takada \cite{24}, the knowledge of $I^{\sigma \sigma'}({\bf p},\epsilon;{\bf p}',\epsilon';{\bf q},\omega)$ 
at ${\bf q} = \omega = 0$ is sufficient for the evaluation of the
ground state energy as well as the self-energy $\Sigma_{\sigma}({\bf p},\epsilon)$.

    It is important to recognize that the expansion of the self-energy
$\Sigma_{\sigma}({\bf p},\epsilon)$ or the proper polarization function $\pi({\bf q},\omega)$ with respect to the
screened Coulomb interaction converges very slowly since correlation is
much complicated in the
region of realistic metallic densities. It may then
be naturally expected that the systematic inclusion of higher-order perturbation terms
with respect to the screened Coulomb interaction is indispensable for the
description of strongly correlated features of the electron liquid. In
fact, there is a great deal of cancellation due to the Pauli principle
occurring among various higher-order perturbation terms. For the evaluation
of the quasi-particle energy dispersion $E({\bf p})$ as well as the effective mass
$m^*$, it is necessary to consider the integral equation for the particle-hole
irreducible interaction which makes it possible to take systematically into
account the cancellation between higher-order terms due to the Pauli
principle.

     Finally we give a brief survey of the present situation in the study
of the electron liquid. (1) The numerical values of the correlation energy
calculated from different theoretical methods \cite{22,23,29,30,31,32,33,34,35,36} 
as well as from the Green's function Monte Carlo method \cite{29,30} agree within an accuracy of 0.5 mRy. per
electron throughout the whole region of metallic densities. The
compressibility and the spin-susceptibility are also evaluated with high
accuracy. The RPA overestimates the correlation energy in
magnitude by about $47\%$ at a typical metallic density $(r_s \sim 4.0)$ since it
lacks higher-order particle-particle ladder interactions responsible for
short-range correlation as well as exchange corrections. (2) Accurate
evaluations of the spin-parallel and the spin-antiparallel pair correlation
functions are performed \cite{37,38,39}. (3) Physical quantities of quasi-particles
such as the effective mass $m^*$ \cite{16,40,41}, the renormalization factor z \cite{42}, the Landau
interaction function $f^{\sigma \sigma'}({\bf p,p'})$ \cite{16} and the quasiparticle energy dispersion
$E({\bf p})$ \cite{43,44,45,46} (see Sec. V) are all evaluated with sufficient accuracy. (4) A
self-consistent theory in which the integral equation for the particle-hole
irreducible interaction can be fulfilled by iteration in the framework of
the conserving approximation scheme has been presented by Takada \cite{47} and
successfully applied to the quantitative evaluation of the one-electron
spectral function $A({\bf p}, \epsilon)$ \cite{48} and the dynamical structure factor $S({\bf q}, \omega)$ \cite{49} of the
electron liquid at metallic densities.
\section{NEW EXPRESSION FOR CORRELATION ENERGY}
\subsection{Definition}
In this section we define a new expression for the correlation energy of
the electron liquid as follows:
\begin{eqnarray}
E_c&=&\frac{1}{2}\left(\frac{1}{\Omega}\right)^2\sum_{\bf q}\sum_{{\bf p},\sigma, {\bf p}',\sigma'}
f({\bf p})f({\bf p'})\nonumber\\
&\times&v({\bf q})\frac{\{1-f({\bf p+q})\}\{1-f({\bf p'-q})\}}
{\epsilon_{p}-\epsilon_{{\bf p+q}}+\epsilon_{p'}-\epsilon_{ {\bf p'-q} }}\nonumber \\
&\times&\left\{v_{eff}({\bf q})-\delta_{\sigma \sigma'}v_{eff}({\bf -p+p'-q})\right\},\label{eq:3.1}
\end{eqnarray}
where $f({\bf p})$ is the Fermi distrubution function at 0K and 
$\epsilon_p=\frac{\hbar^2 p^2}{2m}$. 
The expression bears second-order direct and exchange perturbation terms, but
one of the two Coulomb interactions in each term is replaced by an
effective interaction $v_{eff}({\bf q})$ which contains information about third- and
higher-order contributions in the perturbation expansion. In Sec. III B, C
and D we evaluate the effective interaction $v_{eff}({\bf q})$ by interpolating
between long-range correlation in the RPA and short-range correlation in
the particle-particle ladder approximation such that the exchange
counterpart $v_{eff}({\bf -p+p'-q})$ 
and its feedback effect on the direct interaction
$v_{eff}({\bf q})$ are evaluated in a self-consistent way. This expression for
the correlation energy may be understood from intuition or from a
diagrammatic consideration. Strictly, the effective interaction $v_{eff}({\bf q})$
above should be defined as
$v_{eff}({\bf p, p'; q})$. Here we employ its averaged value over ${\bf p}$ and ${\bf p'}$ within the
Fermi sphere.

     First we shall see how the above expression for $E_c$ is related to the
dielectric formulation we have described in Sec. II B. The direct term of
$E_c$ comes from the direct part of the total polarization function 
$\pi_b({\bf q},\omega)-\pi({\bf q},\omega)v({\bf q}) \pi({\bf q},\omega)/\epsilon({\bf q},\omega)$ 
and the exchange term of $E_c$ from the
exchange part of the total polarization function excluding the zeroth order
Lindhard function $\pi_a({\bf q},\omega)-\pi_0({\bf q},\omega)$.

     We start with the standard expression for the exchange-correlation 
energy, i.e., the integral of the potential energy over the
coupling contant.
\begin{eqnarray}
E_{xc}=\frac{n}{2}\sum_{\bf q}\frac{4\pi e^2}{{\bf q}^2}\{\bar{S}({\bf q})-1\},\label{eq:3.2}
\end{eqnarray}
where $n$ is the electron density and $\bar{S}({\bf q})$ denotes the
coupling-constant-averaged static structure factor defined as
\begin{eqnarray}
\bar{S}({\bf q})=\frac{1}{e^2}\int_0^{e^2}d(e^2)S({\bf q}),~~~S({\bf q})=\frac{1}{N}\langle \rho_{\bf q}\rho_{-\bf q}\rangle.\label{eq:3.3}
\end{eqnarray} 
The integral over the coupling constant necessarily reduces the magnitude
of the potential energy since it implies the addition of an increase in 
the kinetic energy caused by correlation to the potential energy.
This correlational increase in the kinetic energy cannot be neglected since
it cancels about one third of the correlational lowering in the potential
energy at a typical metallic density, $r_s=4.0$; in the limit $r_s \to 0$ the
correlational lowering in the potential energy is by half cancelled by the
correlational increase in the kinetic energy.
 
    The function $\bar{S}({\bf q})$ can be decomposed into two parts as
$\bar{S}({\bf q})=S^{HF}({\bf q})+\bar{S}_c({\bf q})$. The Hartree-Fock contribution $S^{HF}({\bf q})$ is 
independent of the coupling constant. Then the correlation energy is given as
\begin{eqnarray}
E_c=\frac{n}{2}\sum_{\bf q}\frac{4\pi e^2}{{\bf q}^2}\bar{S}_c({\bf q}),\label{eq:3.4}
\end{eqnarray}
where $\bar{S}_c({\bf q})$ can be decomposed into direct and exchange parts,
$\bar{S}({\bf q})={\bar{S}_c}^d({\bf q})+{\bar{S}_c}^{ex}({\bf q})$.

     Let us compare Eq. (\ref{eq:3.4}) with the new expression for $E_c$ we have defined
by Eq. (\ref{eq:3.1}). Then the functions ${\bar{S}_c}^d({\bf q})$ and ${\bar{S}_c}^{ex}({\bf q})$ may be identified as
\begin{eqnarray}
&&{\bar{S}_c}^d({\bf q})=\frac{1}{n}\left(\frac{1}{\Omega}\right)^2\sum_{\bf p,p'}\sum_{\sigma,\sigma'}f({\bf p})f({\bf p}')\nonumber~~~~~~~~~~~~~~~~~\\
&&~~~~~~~\times\frac{\{1-f({\bf p+q})\}\{1-f({\bf p'-q})\}}{\epsilon_{p}-\epsilon_{{\bf p+q}}+\epsilon_{p'}-\epsilon_{{\bf p'-q} }}v_{eff}({\bf q}),\label{eq:3.5}\\
&&{\bar{S}_c}^{ex}({\bf q})=-\frac{1}{n}\left(\frac{1}{\Omega}\right)^2\sum_{\bf p,p'}\sum_{\sigma,\sigma'}f({\bf p})f({\bf p'})\nonumber~~~~~~~~~~~~~~~~~\\
&&~~~~~~~~~~\times\frac{\{1-f({\bf p+q})\}\{1-f({\bf p'-q})\}}{\epsilon_{p}-\epsilon_{{\bf p+q}}+\epsilon_{p'}-\epsilon_{ {\bf p'-q} }}\nonumber\\
&&~~~~~~~~~~\times\delta_{\sigma \sigma'}v_{eff}({\bf -p+p'-q})\label{eq:3.6}.
\end{eqnarray}
The spin-parallel and spin-antiparallel components of ${\bar{S}_c}({\bf q})$ are defined as
${\bar{S}_c}^{d}({\bf q})/2$ and ${\bar{S}_c}^{d}({\bf q})/2+{\bar{S}_c}^{ex}({\bf q})$, respectively, and given as
\begin{eqnarray}
&&{\bar{S}_c}^{\uparrow \downarrow}({\bf q})=\frac{1}{2n}\left(\frac{1}{\Omega}\right)^2\sum_{\bf p,p'}\sum_{\sigma,\sigma'}f({\bf p})f({\bf p}')\nonumber~~~~~~~~~~~~~~~~~\\
&&~~~~~~\times \frac{\{1-f({\bf p+q})\}\{1-f({\bf p'-q})\}}{\epsilon_{p}-\epsilon_{{\bf p+q}}+\epsilon_{p'}-\epsilon_{ {\bf p'-q} }}v_{eff}({\bf q}),\label{eq:3.7}\\
&&{\bar{S}_c}^{\uparrow \uparrow}({\bf q})=\frac{1}{2n}\left(\frac{1}{\Omega}\right)^2\sum_{\bf p,p'}\sum_{\sigma,\sigma'}f({\bf p})f({\bf p}')\nonumber\\
&&~~~~~~~~~~\times\frac{\{1-f({\bf p+q})\}\{1-f({\bf p'-q})\}}{\epsilon_{p}-\epsilon_{{\bf p+q}}+\epsilon_{p'}-\epsilon_{ {\bf p'-q} }}v_{eff}({\bf q})\nonumber\\
&&~~~~~~~~~~-\frac{1}{n}\left(\frac{1}{\Omega}\right)^2\sum_{{\bf p},\sigma, {\bf p}',\sigma'}f({\bf p})f({\bf p}')\nonumber\\
&&~~~~~~~~~~\times\frac{\{1-f({\bf p+q})\}\{1-f({\bf p'-q})\}}
{\epsilon_{p}-\epsilon_{{\bf p+q}}+\epsilon_{p'}-\epsilon_{ {\bf p'-q} }}\nonumber\\
&&~~~~~~~~~~\times\delta_{\sigma \sigma'}v_{eff}({\bf -p+p'-q}).\label{eq:3.8}
\end{eqnarray}
Note that this expression for ${\bar{S}_c}^{\uparrow \uparrow}({\bf q})$ fulfills the
Pauli principle. The coupling-constant-averaged spin-parallel pair
correlation function is defined as
\begin{eqnarray}
\bar{g}^{\uparrow \uparrow}({\bf r})=
1+\frac{2}{N}\sum_{\bf q} \left\{\bar{S}^{\uparrow \uparrow}({\bf q})-1\right\}e^{i{\bf q \cdot r}}.\label{eq:3.9}
\end{eqnarray}
Since $S^{HF}({\bf q})$ satisfies that $\bar{g}^{\uparrow \uparrow}(0)=0$,
the relation that $\sum_{\bf q}{\bar{S}^{\uparrow \uparrow}}_c({\bf q})=0$ has to be satisfied. This identity can
be easily checked by a simple transformation of the wavenumber variables in
Eq. (\ref{eq:3.8}). The Pauli principle is automatically fulfilled in the present
theory. Instead, we have to deal with much difficulty in the
self-consistent determination of the effective interaction $v_{eff}({\bf q})$, as will
be seen in Sec. III B, C and D.
\subsection{New expression for the RPA correlation energy}
In this subsection we calculate the effective interaction $v_{eff}({\bf q})$
appropriate for the RPA. The correlation contribution of the static
structure factor in the RPA is given as
\begin{eqnarray}
S_c^{RPA}({\bf q})=-S^{HF}({\bf q})+\frac{\hbar}{\pi n}\int_0^{\infty}d\omega{\rm Im}\frac{\pi_0({\bf q},\omega)}{1+v({\bf q})\pi_0({\bf q},\omega)}\nonumber,\label{eq:3.10}\\
\end{eqnarray}
where $\pi_0({\bf q},\omega)$ is the Lindhard function. The coupling constant integral of
$S_c^{RPA}({\bf q})$ can be easily performed as
\begin{eqnarray}
\bar{S}_c^{RPA}({\bf q})&=&-S^{HF}({\bf q})\nonumber\\
&+&\frac{\hbar}{\pi n}\int_0^{\infty}d\omega\frac{1}{v({\bf q})}
\tan^{-1}\frac{v({\bf q}){\rm Im}[\pi_0({\bf q},\omega)] }{1+v({\bf q}){\rm Re}[\pi_0({\bf q},\omega)]}.\nonumber\label{eq:3.11}\\
\end{eqnarray}
This expression contains contributions of one-pair excitations and those of
plasmon excitations. The frequency integral in Eq. (\ref{eq:3.11}) can be numerically performed
to obtain $\bar{S}_c^{RPA}({\bf q})$. The effective interaction in the RPA,
$v_{eff}^{RPA}({\bf q})$ is then defined as
\begin{eqnarray}
\bar{S}_c^{RPA}({\bf q})&=&\frac{1}{n}\left(\frac{1}{\Omega}\right)^2\sum_{{\bf p},{\bf p}}\sum_{\sigma,\sigma'}f({\bf p})f({\bf p}')\nonumber\\
&\times&
\frac{\{1-f({\bf p+q})\}\{1-f({\bf p'-q})\}}{\epsilon_{p}-\epsilon_{{\bf p+q}}+\epsilon_{p'}-\epsilon_{ {\bf p'-q} }}v_{eff}^{RPA}({\bf q})\nonumber.\\
\label{eq:3.12}
\end{eqnarray}
The integral over ${\bf p}$ and ${\bf p'}$ in the prefactor has been analytically performed
by Kimball \cite{50}. With the interaction $v_{eff}^{RPA}({\bf q})$ thus calculated we can give a
new expression for the RPA correlation energy as
\begin{eqnarray}
E_c^{RPA}&=&\frac{1}{2}\left(\frac{1}{\Omega}\right)^2\sum_{\bf q,p,p'}\sum_{\sigma,\sigma'}f({\bf p})f({\bf p}')\nonumber\\
&\times& v({\bf q})\frac{\{1-f({\bf p+q})\}\{1-f({\bf p'-q})\}}
{\epsilon_{p}-\epsilon_{{\bf p+q}}+\epsilon_{p'}-\epsilon_{ {\bf p'-q} }}v_{eff}^{RPA}({\bf q})\nonumber.\\
\label{eq:3.13}	
\end{eqnarray}
Note that the exchange term is missing in the RPA. The ratio of the
effective interaction $v_{eff}^{RPA}({\bf q})$ to the bare Coulomb interaction $v({\bf q})$ is
quadratic for small $q$, increases monotonically with ${q}$ and tends to unity
for large ${q}$.

    In Sec. III C and D we employ the expression for $\bar{S}_c^{RPA}({\bf q})$ 
given by Eq.(\ref{eq:3.11}) as a starting point to allow for short-range correlation and exchange.
\subsection{Interpolation between long- and short-range correlations}
As has been mentioned in Sec. III B, the effective interaction evaluated in
the RPA, $v_{eff}^{RPA}({\bf q})$ is reduced in magnitude from the bare Coulomb
interaction $v({\bf q})$ on the small side of wavenumbers under the influence of
dynamical screening in the RPA. In fact, the effective interaction $v_{eff}({\bf q})$
is also reduced in magnitude from $v({\bf q})$ on the large side of wavenumbers
under the strong influence of short-range correlation at metallic
densities; it is not so drastic for small $q$ but essential for large $q$.
\begin{figure}[tb]
  \includegraphics[height=4cm,keepaspectratio]{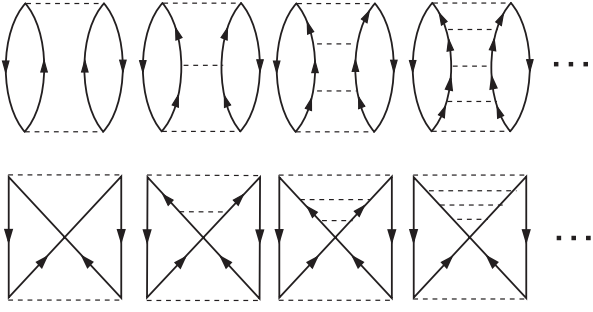}
\caption{\label{fig3} Goldstone energy diagrams with an infinite series of
particle-partricle ladder interactions and their exchange counterparts.}
\end{figure}

Consider those perturbation terms which consist only of an infinite
series of particle-particle ladder interactions \cite{51,52,53,54,55} via the bare Coulomb
interaction. Fig. 3 gives a diagrammatic representation of those perturbation terms. 
The correlation energy contribution from those diagrams is
written as follows:
\begin{eqnarray}
E_c^{ladd (d)}&=&\frac{1}{2}\left(\frac{1}{\Omega}\right)^2\sum_{\bf q,p,p'}\sum_{\sigma,\sigma'}f({\bf p})f({\bf p'})\nonumber\\
&\times&v({\bf q})\frac{ \{1-f({\bf p+q}) \} \{ 1-f({\bf p'-q}) \} }{\epsilon_{p}-\epsilon_{\bf p+q}+\epsilon_{p'}-\epsilon_{\bf p'-q} }I({\bf p,p';q}),\nonumber\\\label{eq:3.14}\\
E_c^{ladd (ex)}&=&-\frac{1}{2}\left(\frac{1}{\Omega}\right)^2\sum_{\bf q,p,p'}\sum_{\sigma,\sigma'}\delta_{\sigma,\sigma'}f({\bf p})f({\bf p'})\nonumber\\
&\times&v({\bf q})\frac{ \{1-f({\bf p+q}) \} \{ 1-f({\bf p'-q}) \} }{\epsilon_{p}-\epsilon_{\bf p+q}+\epsilon_{p'}-\epsilon_{\bf p'-q} }\nonumber\\
&\times&I({\bf p,p';-p+p'-q}),\label{eq:3.15}
\end{eqnarray}
where the particle-particle ladder interaction $I({\bf p,p'; q})$ is the solution
of the following integral equation:
\begin{eqnarray}
&&I({\bf p,p'; q})=v({\bf q})\nonumber\\
&+&\frac{1}{\Omega}\sum_{\bf k}v({\bf q-k})\frac{\{1-f({\bf p+k})\}\{1-f({\bf p'-k})\}}
{\epsilon_{p}-\epsilon_{{\bf p+k}}+\epsilon_{p'}-\epsilon_{ {\bf p'-k} }}\nonumber\\
&\times&I({\bf p,p'; k}).\label{eq:3.16}
\end{eqnarray}
The spin-antiparallel and spin-parallel pair correlation functions,
$g^{\uparrow \downarrow}({\bf r})$ and $g^{\uparrow \uparrow}({\bf r})$ derived from $E_c^{ladd(d)}$ and $E_c^{ladd(ex)}$ are
given respectively as
\begin{eqnarray}
&&g^{\uparrow \downarrow}_{ladd}({\bf r})=\left(\frac{2}{N}\right)^2\sum_{\bf p}\sum_{\bf p'}f({\bf p})f({\bf p}')\nonumber\\
&&\times\left| 1+\frac{1}{\Omega}\sum_{\bf q}\frac{\{1-f({\bf p+q})\} \{1-f({\bf p'-q})\}}
{ \epsilon_{p}-\epsilon_{\bf p+q}+\epsilon_{p'}-\epsilon_{ \bf p'-q }}\right.\nonumber\\
&&~~~~~~~~~~~~~~~~~~~~~~~~~~~~~\times I({\bf p,p'; q})e^{i{\bf q\cdot r}}\biggl|^2,\label{eq:3.17}
\end{eqnarray}
\begin{eqnarray}
&&g^{\uparrow \uparrow}_{HF ladd}({\bf r})=g^{\uparrow \uparrow}_{HF}({\bf r})+\Delta g^{\uparrow \uparrow}_{ladd}({\bf r})\nonumber\\
&=&\left(\frac{2}{N}\right)^2\sum_{\bf p}\sum_{\bf p'}f({\bf p})f({\bf p}')
\frac{1}{2}\left|(1-e^{i({\bf p-p'})\cdot{\bf r} })\right.\nonumber\\
&&\left. + \frac{1}{\Omega}\sum_{\bf q} \frac{\{1-f({\bf p+q})\} \{1-f({\bf p'-q})\}}
{ \epsilon_{p}-\epsilon_{\bf p+q}+\epsilon_{p'}-\epsilon_{ \bf p'-q }}\right.\nonumber\\
&&~~~~~~\times \left.I({\bf p,p'; q})e^{i{\bf q\cdot r}}(1-e^{i({\bf p-p'+2q})\cdot{\bf r} })\right|^2.\label{eq:3.18}
\end{eqnarray}
Eq. (\ref{eq:3.18}) includes the
Hartree-Fock contribution. Both of $g^{\uparrow \downarrow}_{ladd}({\bf r})$ and 
$g^{\uparrow \uparrow}_{HF}({\bf r})+\Delta g^{\uparrow \uparrow}_{ladd}({\bf r})$,
so far as their behaviors at short distances are
concerned, are physically sound and reasonable
throughout the entire region of metallic densities in contrast with the
RPA in which both of the pair correlation functions become negative.

    Consider the general properties of $I({\bf p,p'}; {\bf q})$. It is
reduced to the bare Coulomb interaction $v({\bf q})$ in the limit of small $q$. On the other hand,
it tends approximately to $v({\bf q})[g_{ladd}^{\uparrow \downarrow}(0)]^{1/2}$ in the limit of large $q$.
The ratio $I({\bf p, p'; q})/v({\bf q})$  starts from unity at $q=0$ and decreases monotonically with
increasing $q$.  Thus, $I({\bf p, p'; q})$ is reduced in magnitude
from $v({\bf q})$ on the large side of wavenumbers under the strong influence of
short-range correlation at metallic densities. The inclusion of an infinite series of
particle-particle ladder interactions has the effect to reduce the
contribution from large wavenumber components of one of the two Coulomb
interactions entering the second-order direct perturbation term. 

A very good analytic expression for $g_{ladd}^{\uparrow \downarrow}({\bf r})$ at short distances was given by
one of the authors (H.Y.) \cite{51}. The magnitude of $g^{\uparrow \downarrow}_{ladd}(0)$ is remarkably
reduced from unity as the electron density is lowered to metallic levels. 
The inclusion of particle-particle ladder interactions much reduces 
the ratio $v_{eff}({\bf q})/v({\bf q})$ from unity on the large side of
wavenumbers, compared with the case of the RPA.

     It should be indicated that among all many-body
perturbation terms only particle-particle ladder interacting parts (via
the bare Coulomb interaction) can make a contribution to the predominant
asymptotic form of $S^{\uparrow \downarrow}_c({\bf q})$ of order $q^{-4}$ for large $q$. The predominant
asymptotic form of $S^{\uparrow \uparrow}_c({\bf q})$, on the
other hand, is of order $q^{-6}$ for large
$q$ as a consequence of the cancellation between a direct and exchange pair of 
particle-particle ladder interactions.

     Consider the iterative solution of the integral equation for 
$I({\bf p, p'; q})$. The $n$-th iterative term
involves $n+1$ different wavenumber components of the Coulomb interactions.
Each wavenumber can vary in magnitude from zero to infinity under the
condition that the total sum of $n+1$ different wavenumbers amounts to ${\bf q}$. In
this sense the mixing of different wavenumber components in $I({\bf p, p'; q})$ is
most complete among all types of interactions. It is this property of the
particle-particle ladder interactions that is responsible for the description of
local density fluctuations. The particle-particle ladder approximation is
in marked contrast with the RPA appropriate for the description of
long-range correlation associated with small wavenumber components of the Coulomb interaction. 
The RPA amounts to an infinite summation of the most divergent terms of each order in the
perturbation expansion and repeats infinite times the same wavenumber component ${\bf q}$ of the
Coulomb interaction instead of  mixing different wave number components.
  
   Let us approximate $I({\bf p, p'; q})$ as $v({\bf q})(1-C({\bf q}))$ by averaging over $\bf p$ and
$\bf p'$ within
the Fermi sphere. The local field factor from short-range Coulomb repulsion
$C({\bf q})$ thus defined is quadratic for small $q$, increases monotonically with $q$
and tends approximately to a constant value of $1-[g^{\uparrow \downarrow}_{ladd}(0)]^{1/2}$.
   
  We make a  smooth interpolation between long-range correlation
in the RPA and short-range correlation in the particle-partricle ladder
approximation by replacing every bare Coulomb interaction $v({\bf q})$ appearing in
Eq.(\ref{eq:3.11}) by $v({\bf q})(1-C({\bf q}))$ as
\begin{eqnarray}
&&\bar{S}_c^{RPA+ladd(d)}({\bf q})=-S^{HF}({\bf q})+\frac{\hbar}{\pi n}\int_0^{\infty}d\omega\frac{1}{v({\bf q})\{1-C({\bf q})\}}\nonumber\\
&&~~~~~~~~~~\times \tan^{-1}\frac{v({\bf q})(1-C({\bf q})){\rm Im}[\pi_0({\bf q},\omega)] }{1+v({\bf q})(1-C({\bf q})){\rm Re}[\pi_0({\bf q},\omega)}.\label{eq:3.19}
\end{eqnarray}
It is evident from the interpolation that this expression is reduced
to $\bar{S}_c^{RPA}({\bf q})$ for small $q$ and tends to
$\bar{S}_c^{ladd(d)}({\bf q})$ for large $q$. Following
the same procedure as in the RPA, we have evaluated the
effective interaction appropriate for the interpolation. The ratio
of the interpolated effective interaction ${v}_{eff}^{RPA+ladd(d)}({\bf q})$ to $v({\bf q})$ is
quadratic for small $q$ in a similar way as in the RPA, increases
monotonically with $q$ and tends to a constant of $[g^{\uparrow \downarrow}_{ladd}(0)]^{1/2}$ which
is remarkably smaller than unity at metallic densities. The value of $[g^{\uparrow \downarrow}_{ladd}(0)]^{1/2}$ is
estimated to be $0.547$ at $r_s=2$ and $0.326$ at $r_s=4$, in the
particle-particle ladder approximation \cite{51}.

The exchange counterpart of particle-particle ladder interactions shown in Fig. 3 will
be considered in Sec. III D, combined with other exchange counterparts
associated with small and intermediate wavenumbers.
\subsection{Allowance for the exchange correction}
In this subsection we allow for the exchange correction, starting
with the interpolated effective interaction we have obtained above. First
of all, we would like to remind the reader of the detailed analysis of the
dielectric formulation we have made in terms of the exchange and direct
parts of the proper polarization function, $\pi_a({\bf q},\omega)$ and $\pi_b({\bf q},\omega)$ in Sec. II.B. 
So far we have treated the Lindhard function $\pi_0({\bf q},\omega)$ that is the non-interacting contribution of
$\pi_a({\bf q},\omega)$, the main contribution of the direct part
$\pi_b({\bf q},\omega)$ involving an infinite series of particle-particle 
ladder interactions and the corresponding improper polarization
function, leaving the interacting exchange part $\pi_a({\bf q},\omega)-\pi_0({\bf q},\omega)$ out of
account. In Sec. II. B we have stressed that the exchange part $\pi_a({\bf q},\omega)$ has
an influence on
$S^{\uparrow \uparrow}({\bf q})$ and
$S^{\uparrow \downarrow}({\bf q})$ indirectly through the improper polarization function common
to the two. This means that the two functions $S^{\uparrow \uparrow}({\bf q})$ and
$S^{\uparrow \downarrow}({\bf q})$
are interconnected through $\pi_a({\bf q},\omega)$ in the improper polarization function
and have to be determined in a self-consistent way. For convenience we term
the requirement of this self-consistency through the improper polarization
function the indirect, or feedback effect of exchange.
   
  In order to allow for the indirect effect of exchange as well as the
direct one  we employ an iterative method. First, we start with an exchange
interaction of ${v_{eff}}^{RPA+ladd(d)}({\bf -p+p'-q})$, which is 
obtained by replacing ${\bf q}$ in ${v_{eff}}^{RPA+ladd(d)}({\bf q})$ by the variable $\bf -p+p'-q$
appropriate for the exchange term. Secondly, we define the local field
factor from exchange $G(q)$ as follows:
\begin{eqnarray}
&&-\frac{1}{n}\left(\frac{1}{\Omega}\right)^2\sum_{{\bf p},\sigma, {\bf p}',\sigma'}f({\bf p})f({\bf p}')\delta_{\sigma \sigma '}\nonumber\\
&&~~\times\frac{\{1-f({\bf p+q})\}\{1-f({\bf p'-q})\}}{\epsilon_{p}-\epsilon_{{\bf p+q}}+\epsilon_{p'}-\epsilon_{ {\bf p'-q} }}v_{eff}({\bf-p+p'-q})\nonumber\\
&&=-\frac{1}{n}\left(\frac{1}{\Omega}\right)^2\sum_{{\bf p},\sigma, {\bf p}',\sigma'}f({\bf p})f({\bf p}')\delta_{\sigma \sigma'}\nonumber\\
&&~~~~~~~~~~\times\frac{\{1-f({\bf p+q})\}\{1-f({\bf p'-q})\}}
{\epsilon_{p}-\epsilon_{{\bf p+q}}+\epsilon_{p'}-\epsilon_{ {\bf p'-q} }}v({\bf q})G({\bf q}).\nonumber\\\label{eq:3.20}
\end{eqnarray}
This is the first value of $\bar{S}_c^{ex}({\bf q})$ to be used in the iterative process.
To make the numerical calculation of the left-hand side much easier we have employed
the same transformation of the variables that Geldart and Taylor \cite{56} made in
their study of the electron liquid.

     Thirdly, we replace every factor of $v({\bf q})((1-C({\bf q}))$ appearing in
Eq.(\ref{eq:3.19}) by $v({\bf q})( 1-C({\bf q})-G({\bf q})/2 )$ to allow for the feedback effect of
exchange; a factor of $1/2$ is to be attached because the local field
correction from exchange occurs between electrons with the same spin orientations alone.
The resulting expression is given as
\begin{eqnarray}
&&{\bar{S}_c}^{RPA+ladd(d)+ex}({\bf q})=-S^{HF}({\bf q})\nonumber\\
&&+\frac{\hbar}{\pi n}\int_0^{\infty}d\omega\frac{1}{v(q)(1-C(q)-G(q)/2)}\nonumber\\
&&\times\tan^{-1}\frac{v({\bf q})(1-C({\bf q})-G({\bf q})/2){\rm Im}[\pi_0({\bf q},\omega)] }{1+v({\bf q})(1-C({\bf q})-G({\bf q})/2){\rm Re}[\pi_0({\bf q},\omega)]}.\nonumber\\
\label{eq:3.21}
\end{eqnarray}
Fourthly, we evaluate a new effective interaction $v_{eff}^{NEW}({\bf q})$ from the
direct part of $\bar{S}_c^{RPA+ladd(d)+ex}({\bf q})$, i.e.,
$\bar{S}_c^{RPA+ladd(d)+ex}({\bf q})-{\bar{S}_c}^{ex}({\bf q})$ as
\begin{eqnarray}
&&\bar{S}_c^{RPA+ladd(d)+ex}({\bf q})-{\bar{S}_c}^{ex}({\bf q})\nonumber\\
&&~~~~~~~~~~=\frac{1}{n}\left(\frac{1}{\Omega}\right)^2\sum_{{\bf p},\sigma,{\bf p}',\sigma'}f({\bf p})f({\bf p}')\nonumber\\
&&~~~~~~~~~~\times\frac{\{1-f({\bf p+q})\}\{1-f({\bf p'-q})\}}
{\epsilon_{p}-\epsilon_{{\bf p+q}}+\epsilon_{p'}-\epsilon_{ {\bf p'-q} }}v_{eff}^{NEW}({\bf q}).\nonumber\\ \label{eq:3.22}
\end{eqnarray}
The interaction $v_{eff}^{NEW}({\bf q})$ thus defined includes the feedback effect of
exchange. From $v_{eff}^{NEW}({\bf q})$ we construct the exchange
interaction $v_{eff}^{NEW}({\bf -p+p'-q})$ to calculate a new value of the local field
factor from exchange
$G({\bf q})$. We have repeated the iterative process above until we reach the
convergent result of the local field factor $G({\bf q})$ and the effective
interaction $v_{eff}({\bf q})$.
  
It should be noted that the above self-consistent determination of the
exchange correction corresponds to seeking an approximate solution to the
integral equation for the particle-hole irreducible interaction within the present
form of the correlation energy.
\begin{figure}[tb]
  \includegraphics[width=7.8cm,keepaspectratio]{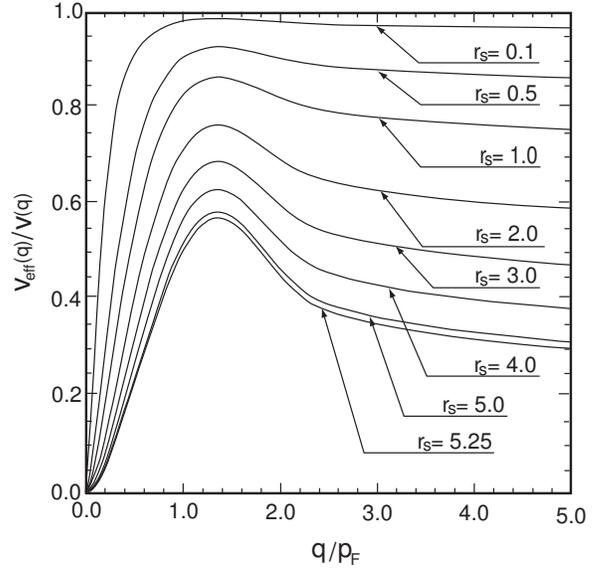}
\caption{\label{fig4} The ratio of the effective interaction to the bare Coulomb
interaction, $v_{eff}({\bf q})/v({\bf q})$ as a function of $q/p_F$ for various values of $r_s$.}
\end{figure}

     The ratio of the self-consistently determined effective interaction
$v_{eff}({\bf q})$ to the bare Coulomb interaction $v({\bf q})$ is shown for various values of
$r_s$ in Fig. 4.
A remarkable change in the ratio $v_{eff}({\bf q})/v({\bf q})$ from its interpolated value
in Sec. III C can be seen over a wide range of intermediate wavenumbers.
The feedback effect of exchange makes the ratio $v_{eff}({\bf q})/v({\bf q})$ have a broad
peak around $q/p_F =1.4$, though its limiting value for large $q$ remains
unchanged.
     
The difference of the ratio $v_{eff}({\bf q})/v({\bf q})$ from unity can be interpreted
to be a measure of how third- and higher-order perturbation terms
effectively reduce in magnitude one of the two Coulomb interactions
entering the second-order direct perturbation term. The feedback effect of
exchange enhances the effective interaction over a wide range of
intermediate wavenumbers, particularly around
$q/p_F =1.4$ almost independently of $r_s$ throughout the entire region of
metallic densities. The origin of the broad peak around $q/p_F =1.4$ can be
traced back to the interacting exchange part $\pi_a({\bf q},\omega)-\pi_0({\bf q},\omega)$. 
This can easily be seen from the starting expression for the function $S^{\uparrow \downarrow}({\bf q})$ in Sec.
II B.
\begin{eqnarray}
       S^{\uparrow \downarrow}({\bf q}) &=& \frac{\hbar}{2\pi n}\int_0^{\infty}{\rm Im}
\left\{\frac{1}{v({\bf q})}\left(1-\frac{1}{\epsilon({\bf q},\omega)}\right)-\pi_a({\bf q},\omega)\right\}\nonumber\\
&=&\frac{1}{2}S_c({\bf q})-\frac{\hbar}{2\pi n}\int_0^{\infty}d\omega{\rm Im}(\pi_a({\bf q},\omega)-\pi_0({\bf q},\omega))\nonumber\\
&=&\frac{1}{2}(S_c({\bf q})-\tilde{S}_c({\bf q})).\label{eq:3.23}
\end{eqnarray}
The broad peak in the ratio $v_{eff}({\bf q})/v({\bf q})$ comes from the peak of $\tilde{S}_c({\bf q})$
or $\bar{\tilde{S}}_c({\bf q})$ located at the almost same wavenumber.
\subsection{Coupling-constant-averaged spin-parallel and spin-antiparallel pair correlation functions}
An exact expression for the exchange and correlation energy functional in
DFT is given in terms of the coupling-constant-averaged pair correlation
function $\bar{g}({\bf r,r'})$ \cite{57,58,59,60} whose behaviors are analogous to those of the usual
spin-averaged pair correlation function $g({\bf r,r'})$.
\begin{eqnarray}
E_{xc}=\frac{1}{2}\int d{\bf r}d{\bf r'}\frac{n({\bf r})n({\bf r}')}{|{\bf r-r'}|}\{\bar{g}({\bf r},{\bf r}')-1\}.\label{eq:3.24}
\end{eqnarray}
This is the inhomogeneous version of the formula for the exchange and
correlation energy of the uniform electron liquid. Similarly, the integral
over the coupling constant implies the addition of the correlational
increase in the kinetic energy to the correlational lowering in the
potential energy between electrons. The function $\bar{g}({\bf r,r'})$ is defined such that the presence
of a coupling-constant dependent fictitious one-electron potential
introduced in a starting reference Hamiltonian maintains the electron
density $n({\bf r})$ of the real interacting system while the Coulomb interaction
among electrons is adiabatically switched on as a perturbation. This
artifice reflects that correlation among electrons in inhomogeneous systems
necessarily causes a change in the electron density $n({\bf r})$ in contrast with
the uniform electron liquid.

     The magnitude of the pair correlation function $g({\bf r,r'})$
at short separations depends on whether one-electron orbitals concerned are
extended or localized. Probably, the lowering of $g({\bf r,r'})$ from $1/2$ at short
separations will be greater for nearly free electrons than for atomic
electrons tightly bound to the nucleus. This property will also apply to
the function $\bar{g}({\bf r,r'})$.

     In this subsection we shall first discuss what effect the correlation
energy functional generally has on the electron density $n({\bf r})$ of non-uniform
electron systems with the help of the expression for $E_{xc}$ above. The sum of
the Hartree energy functional and the above $E_{xc}$  gives the total
electron-electron interaction energy functional in the ground state energy
functional in DFT. The general properties of $\bar{g}({\bf r,r'})$ are quite analogous
to those of the usual spin-averaged pair correlation function $g({\bf r,r'})$,
though the quantity $\bar{g}({\bf r,r'})-1$ is somewhat reduced in magnitude under the
influence of the correlational increase in the kinetic energy.

     In the Hartree-Fock approximation, $\bar{g}({\bf r,r})=1/2$ and
$1/2<\bar{g}({\bf r,r'})<1$.
Generally, correlation reduces the value of $\bar{g}({\bf r,r'})$ on the
short-distance side of $|{\bf r-r'}|$ and reversely enhances the value of
$\bar{g}({\bf r,r'})$ on the long-distance side such that the total electron charge
involved is conserved; at zero separation $0<\bar{g}({\bf r,r})<1/2$
and the function
$\bar{g}({\bf r,r'})$ increases with $|{\bf r-r'}|$, exceeds
unity and
forms a broad peak around the average inter-electron separation and
thereafter asymptotically approaches to unity. The function $\bar{g}({\bf r,r'})$ can
be rewritten as
$\bar{g}({\bf \bar{r},R})$ with $\bf{\bar{r}=r-r'}$ and ${\bf R=(r+r')}/2$ and its value at zero separation
$\bar{g}(0, {\bf R})$ should vary rather slowly from place to place in solids.
 
    Repelling of other electrons from the immediate neighborhood of an
electron in question, when it approaches each nucleus in solids, makes it
"feel" the attractive potential from the nucleus significantly stronger 
because of less screening than in the Hartree-Fock approximaton in DFT. Hence 
the electron density $n({\bf r})$ is enhanced in the immediate
neighborhood of the nuclei. Generally, correlation causes an enhancement in 
the electron density $n({\bf r})$ at the positions of individual nuclei in solids.

     Next we shall mention the general properties of the
coupling-constant-averaged spin-parallel and spin-antiparallel pair
correlation functions for the uniform electron liquid, $\bar{g}^{\uparrow \uparrow}({\bf r})$ and 
$\bar{g}^{\uparrow \downarrow}({\bf r})$ and their Fourier
transforms $\bar{S}^{\uparrow \downarrow}({\bf q})$ and $\bar{S}^{\uparrow \downarrow}({\bf q})$ 
before we show their numerical results calculated from the present theory. The
coupling-constant-averaged pair correlation functions $\bar{g}^{\sigma \sigma'}({\bf r})$ are defined
as
\begin{eqnarray}
\bar{g}^{\sigma \sigma'}({\bf r})=\frac{1}{e^2}\int_0^{e^2}d(e^2) g^{\sigma \sigma'}({\bf r}).\label{eq:3.25}
\end{eqnarray}
The integral over the coupling constant generally reduces both the shift of $\bar{g}^{\uparrow \uparrow}({\bf r})$
from unity and the shift of $\bar{g}^{\uparrow \downarrow}({\bf r})$ from its Hatree-Fock value
over all distances since the correlational increase in the kinetic energy
is involved. The reduction is conspicuous particularly at short distances.
The cusp condition on $\bar{g}^{\sigma \sigma'}({\bf r})$ \cite{53,55,61,62,63,64,65} is modified as
\begin{eqnarray}
&&\bar{g}^{\uparrow \downarrow}({\bf r})=\left\{1+\frac{r}{a_0}\left(1-\int_0^{e^2}d(e^2)e^2\right)\right\}\bar{g}^{\uparrow \downarrow}(0)+\cdots,\nonumber\\
\label{eq:3.26}\\
&&\bar{g}^{\uparrow \uparrow}({\bf r})=\frac{r^2}{2}\left\{1+\frac{r}{2a_0}\left(1-\frac{1}{(e^2)^2}\int_0^{e^2}d(e^2)e^2\right)\right\}\nonumber\\
&&~~~~~~~~~~~\times\left.\frac{d^2\bar{g}^{\uparrow \uparrow}(r)}{dr^2}\right|_{r=0}+\cdots. \label{eq:3.27}
\end{eqnarray}
The difference between $g^{\sigma \sigma'}({\bf r})$ and $\bar{g}^{\sigma \sigma'}({\bf r})$
in the cusp condition is the presence of the coupling constant
integral operators in the case of $\bar{g}^{\sigma \sigma'}({\bf r})$. 
The cusp conditions on $g^{\uparrow \downarrow}({\bf r})$ and
$g^{\uparrow \uparrow}({\bf r})$ are fulfilled in the particle-particle
ladder approximation and in the Hartree-Fock plus paricle-particle ladder
approximation, respectively. We have already given the expression for
$g^{\uparrow \downarrow}({\bf r})$ and
$g^{\uparrow \uparrow}({\bf r})$ in these approximations in Eqs.(\ref{eq:3.17}) and (\ref{eq:3.18}),
respectively. These cusp conditions hold also for inhomogeneous electron
systems if one averages the value of $g^{\sigma \sigma'}({\bf r,r'})$
on the sphere centered at ${\bf R}=({\bf r+r'})/2$ with a small radius
${\bf|r-r'|}$ \cite{66}. 
\begin{figure}[t]
  \includegraphics[width=7.5cm,keepaspectratio]{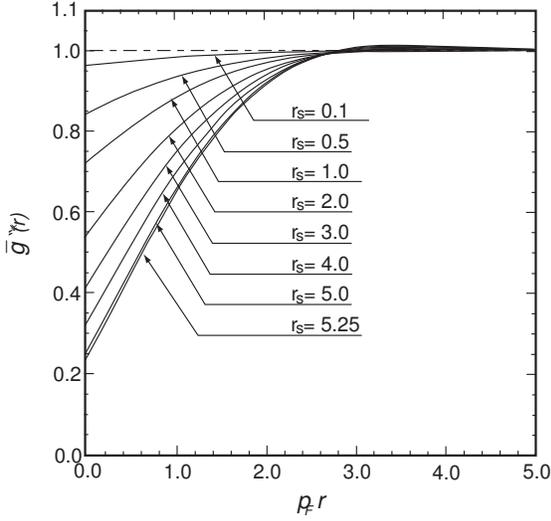}
\caption{\label{fig5} The coupling-constant-averaged spin-antiparallel pair correlatrion
function for the electron liquid, $\bar{g}^{\uparrow \downarrow}({\bf r})$ calculated from the
present theory as a function of $p_Fr$ for various values of $r_s$.}
\end{figure}
\begin{figure}[t]
  \includegraphics[width=7.5cm,keepaspectratio]{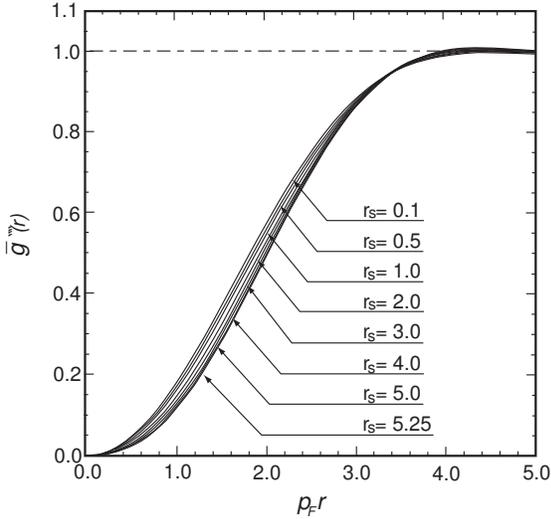}
\caption{\label{fig6} The coupling-constant-averaged spin-parallel pair correlatrion
function for the electron liquid, $\bar{g}^{\uparrow \uparrow}({\bf r})$ calculated from the present
theory as a function of $p_Fr$ for various values of $r_s$.}
\end{figure}

     In Fig. 5 the value of $\bar{g}^{\uparrow \downarrow}({\bf r})$ 
calculated from the present theory is drawn for various values of $r_s$. 
We have ascertained that the value of $\bar{g}^{\uparrow \downarrow}({\bf r})$ is
significantly enhanced at short distances, compared with the value of 
$g^{\uparrow \downarrow}({\bf r})$ calculated in
the particle-particle ladder approximation. As is obvious from Fig. 6, the
present $\bar{g}^{\uparrow \uparrow}({\bf r})$ fulfills the Pauli principle. We
have also observed that the shift of ${\bar g}^{\uparrow \uparrow}({\bf r})$ from its Hartree-Fock
value is reduced at short and intermediate distances, compared with
the case of $g^{\uparrow \uparrow}({\bf r})$ calculated in the same approximation. 
Perdew and Wang \cite{67} have also calculated the spin-averaged 
$\bar{g}(\bf{r})$ of the electron liquid by interpolating between 
the RPA and the particle-particle ladder
approximation, but they have left the interacting exchange part
$\pi_a({\bf q},\omega)-\pi_0({\bf q},\omega)$ out of account.
\begin{figure}[tb]
  \includegraphics[width=7.5cm,keepaspectratio]{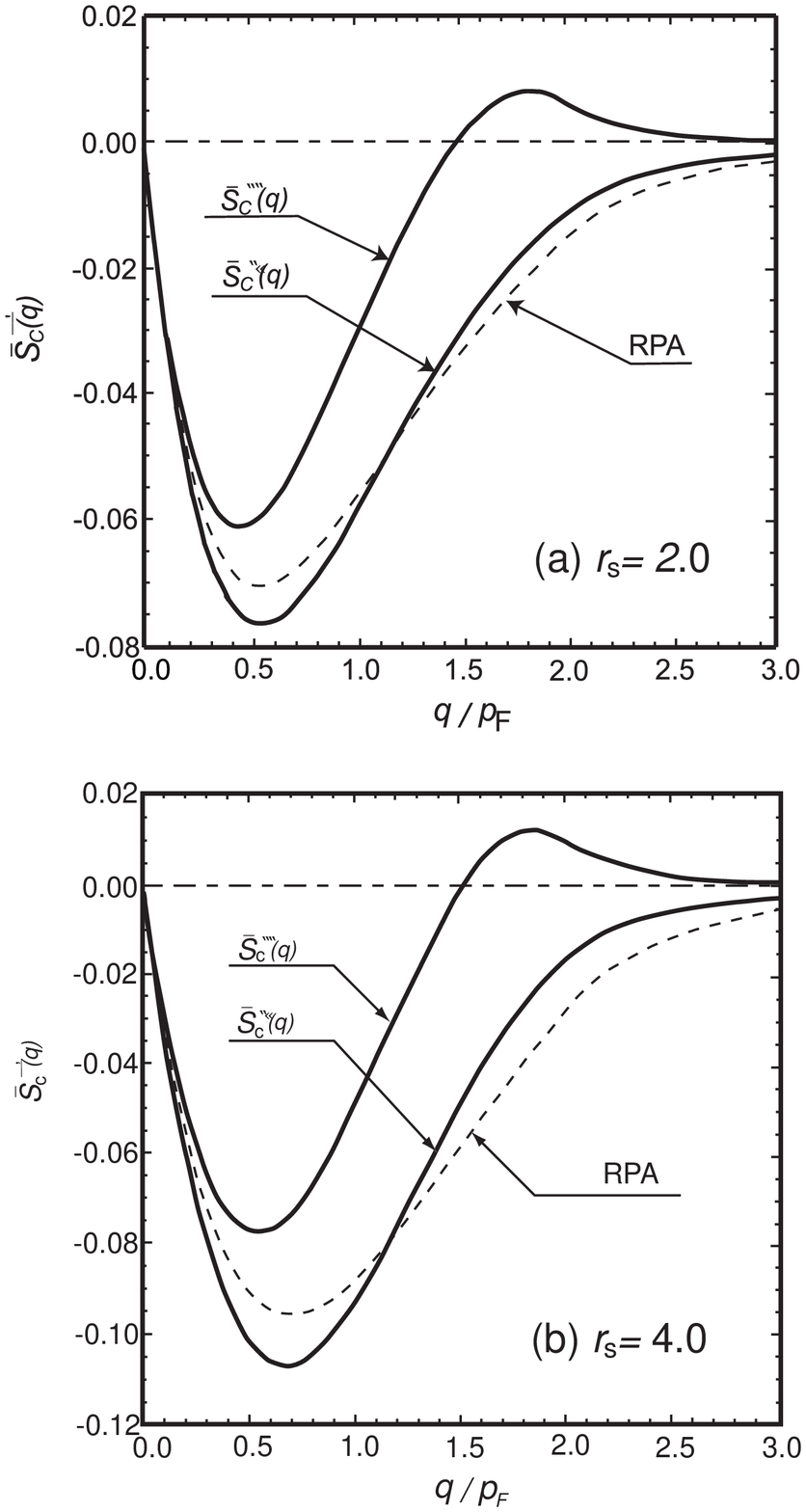}
\caption{\label{fig7} (a) and (b) The coupling-constant-averaged spin-parallel and
spin-antiparallel correlational contributions of the static structure factor,
$\bar{S}_c^{\uparrow \uparrow}({\bf q})$ and ${\bar{S}_c}^{\uparrow \downarrow}({\bf q})$
as functions of $q/p_F$ for $r_s=2.0$ and $4.0$; the dotted line in each figure represents 
the value of the two functions in the RPA.}
\end{figure}

     Fig. 7 shows the functions $\bar{S}_c^{\uparrow \uparrow}({\bf q})$ and
$\bar{S}_c^{\uparrow \downarrow}({\bf q})$ in the
present theory. For comparison the same functions in the RPA are also drawn
in the figure. These functions are both reduced in magnitude from the
corresponding functions without bars for the same reason as in the case of
$\bar{g}^{\sigma \sigma'}({\bf r})$. The difference between $\bar{S}_c^{\uparrow \uparrow}({\bf q})$ and
$\bar{S}_c^{\uparrow \downarrow}({\bf q})$ is given
by  $\bar{\tilde{S}}_c({\bf q})$. For lack of the second- and higher-order exchange
corrections the RPA fails to discriminate between the two functions. As is
obvious from Fig. 7, the inclusion of exchange corrections shifts the two
functions in opposite directions for small $q$. This splitting is evident
from the very definition of $S^{\uparrow \uparrow}({\bf q})$ and
$S^{\uparrow \downarrow}({\bf q})$ that we have given in Sec. II B. 
The interacting exchange part of proper polarization function
$\pi_a({\bf q},\omega)-\pi_0({\bf q},\omega)$ is responsible for this splitting of 
$\bar{S}_c^{\uparrow \uparrow}({\bf q})$
and $\bar{S}_c^{\uparrow \downarrow}({\bf q})$ for small $q$. 
In this connection we mention that the accurate
description of $S({\bf q})$ in the RPA for small $q$ does not imply that it also
can give the accurate description of $\tilde{S}({\bf q})$ for small $q$.

    The function $\bar{S}_c^{\uparrow \uparrow}({\bf q})$ becomes positive for $q \gtrsim 1.5p_F$ while the
function $\bar{S}_c^{\uparrow \downarrow}({\bf q})$ stays negative for any $q$. This is because the
exchange contribution overcomes in magnitude the direct contribution for
intermediate and large wavenumbers. The following asymptotic forms of
${S}_c^{\uparrow \downarrow}({\bf q})$ and
${S}_c^{\uparrow \uparrow}({\bf q})$ for large $q$ are an alternative representation of
the cusp conditions \cite{55} on $g^{\uparrow \downarrow}({\bf r})$ and
$g^{\uparrow \uparrow}({\bf r})$. 
\begin{eqnarray}
{S}_c^{\uparrow \downarrow}({\bf q})=-\frac{4}{3}\left(\frac{\alpha r_s}{\pi}\right)\left(\frac{p_F}{q}\right)^4g^{\uparrow \downarrow}(0)+\cdots,\label{eq:3.28}\\
{S}_c^{\uparrow \uparrow}({\bf q})=4\left(\frac{\alpha r_s}{\pi}\right)\left(\frac{p_F}{q}\right)^6\frac{d^2g^{\uparrow \uparrow}(r)}{d(p_Fr)^2}+\cdots.\label{eq:3.29}
\end{eqnarray}
It is straightfoward to derive the corresponding asymptotic forms of
$\bar{S}_c^{\uparrow \downarrow}({\bf q})$
and $\bar{S}_c^{\uparrow \uparrow}({\bf q})$ for
large $q$.

     In Figs. 8 and 9 ${\bar{S}_c}^{\uparrow \downarrow}({\bf q})$
and ${\bar{S}_c}^{\uparrow \uparrow}({\bf q})$ are
drawn for various values of $r_s$, respectively. The correlation energy of the
electron liquid can be evaluated by integrating ${\bar{S}_c}^{\uparrow \downarrow}({\bf q})$ and
$\bar{S}_c^{\uparrow \uparrow}({\bf q})$ over $q$, or equivalently by the area of the two curves drawn
in Figs. 8 and 9 relative to the horizontal axis. The value of the
correlation energy calculated from the present theory agrees with the most
reliable values within an accuracy of 0.5 mRy. per electron throughout the
entire region of metallic densities (see Table I). However, one must not
forget that the accuracy of the correlation energy is not necessarily the
best criterion for the validity of a theory. For example, the
self-consistent GW approximation gives a very accurate evaluation of the
correlation energy of the electron liquid, but it makes the value of 
$g({\bf r})$ negative at short distances \cite{68}.
  
   The present theory gives an accurate description of both $\bar{S}_c^{\uparrow \downarrow}({\bf q})$
and ${\bar{S}_c}^{\uparrow \uparrow}({\bf q})$, or equivalently their Fourier transforms. 
Furthermore, it gives with high accuracy spin-parallel
and spin-antiparallel components of the correlation energy $\epsilon_c$ over
the entire region of metallic densities. Fig. 10 shows the ratios of
${\epsilon_c}^{\uparrow \downarrow}/\epsilon_c$ and 
${\epsilon_c}^{\uparrow \uparrow}/\epsilon_c$ as functions of $r_s$. In the high density region,
the two ratios are analytically given as
\begin{eqnarray}
&&\frac{{\epsilon_c}^{\uparrow \downarrow}}{\epsilon_c}=\frac{1}{2}-\frac{1}{2}\frac{c_{2x}}{0.0622 {\ln}(r_s)}+\cdots,\nonumber\\
&&\frac{{\epsilon_c}^{\uparrow \uparrow}}{\epsilon_c}=\frac{1}{2}+\frac{1}{2}\frac{c_{2x}}{0.0622\ln(r_s)}+\cdots,\nonumber\\
&&\epsilon_c=0.0622\ln(r_s)+c_{RPA}+c_{2x}+\cdots {\rm Ry.},\nonumber\\
&&c_{RPA}=-0.1422,~~c_{2x}=0.04836.\label{eq:3.30}
\end{eqnarray}
\begin{figure}[tb]
  \includegraphics[width=7.5cm,keepaspectratio]{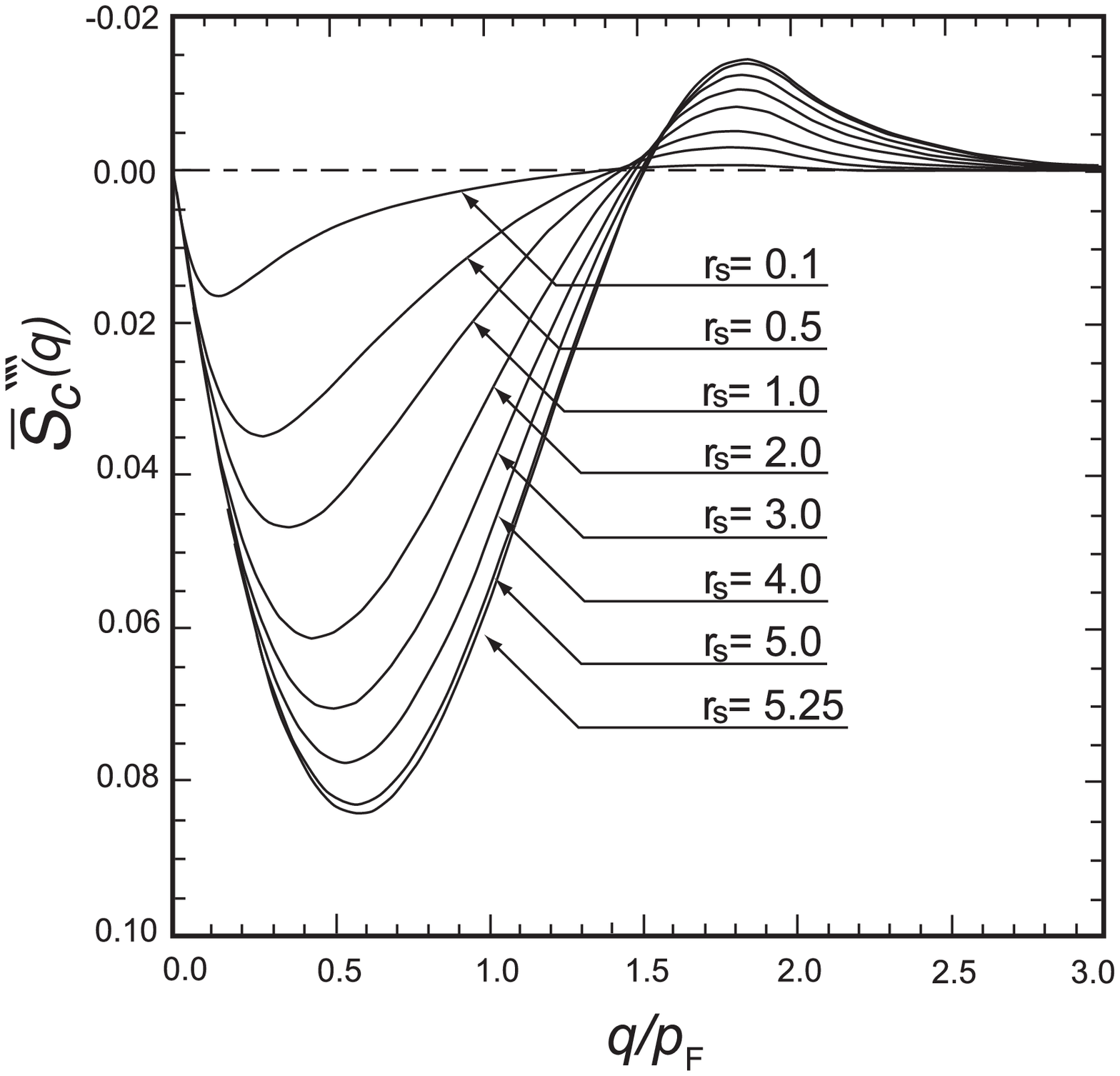}
\caption{\label{fig8} The function ${\bar{S}_c}^{\uparrow \uparrow}({\bf q})$
for various values of $r_s$.}
\end{figure}
\begin{figure}[tb]
  \includegraphics[width=7.4cm,keepaspectratio]{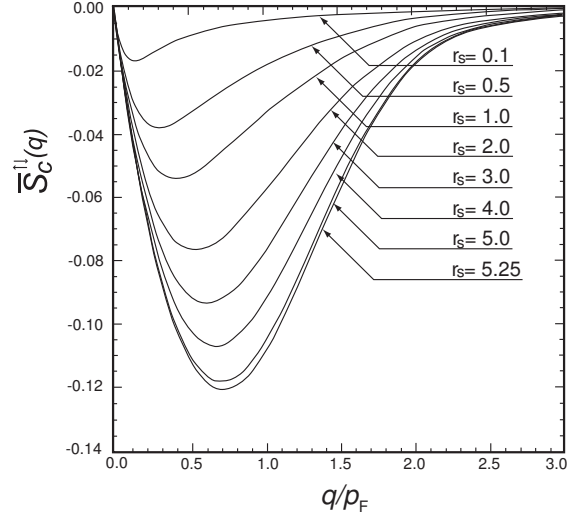}
\caption{\label{fig9} The function ${\bar{S}_c}^{\uparrow \downarrow}({\bf q})$ 
for various values of $r_s$.}
\end{figure}
\hspace*{-3pt}where $c_{RPA}$ is the constant from the RPA and $c_{2x}$ is the value of the
second-order exchange energy \cite{69}. It is important to recognize that $\epsilon_c^{\uparrow \downarrow}$
and $\epsilon_c^{\uparrow \uparrow}$ occupy nearly $70\%$ and $30\%$ of the total correlation energy
throughout almost the entire region of metallic densities, respectively; 
these ratios probably will apply to all valence electrons participating 
in cohesion of solids.
The component $\epsilon_c^{\uparrow \uparrow}$ of $\epsilon_c$ comes from the deepened Fermi hole at short and
intermediate distances from its Hartree-Fock value, as can be seen from
$\bar{g}^{\uparrow \uparrow}({\bf r})$. The component $\epsilon_c^{\uparrow \downarrow}$, on the other hand, comes from the
developed Coulomb hole. The parallel and antiparallel components equivalently contribute
to the leading logarithmic term in the high density limit. 
This is because ${\bar{S}_c}^{\uparrow \downarrow}({\bf q})$ and
${\bar{S}_c}^{\uparrow \uparrow}({\bf q})$ become
equivalent in the limit $r_s \to 0$. This tendency can be observed from the two functions drawn in
Figs. 8 and 9 for $r_s=0.1$. In the language of the pair correlation
functions, the leading
logarithmic term comes from such distances as $1 \ll p_Fr \ll r_s^{-1/2}$ where
$g^{\uparrow \downarrow}({\bf r})-1$ and $g^{\uparrow \uparrow}({\bf r})-g^{\uparrow \uparrow}_{HF}(r)$ both behave like $-r_s/(p_Fr)^2$;
the screening length is proportional to $r_s^{-1/2}/p_F$ for small $r_s$ \cite{55}.
\begin{figure}[t]
  \includegraphics[width=7.5cm,keepaspectratio]{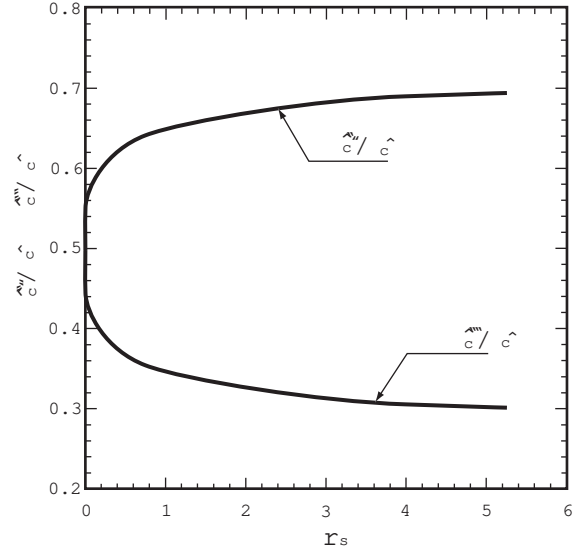}
\caption{\label{fig10} The ratios of $\epsilon_c^{\uparrow \downarrow}/\epsilon_c$
 and $\epsilon_c^{\uparrow \uparrow}/\epsilon_c$ as functions of $r_s$.}
\end{figure}
\begin{table}
\begin{center}
	\begin{tabular*}{8.5cm}{@{\extracolsep{\fill}}c|cccccc}\hline \hline
	 $r_s$ &   RPA  &   STLS   &  C-A   &    VWN  &   EPX  & Present\\ \hline 
	 1.0   & -157.6 &   -123   & -119.6 & -120.6  &  -119  & -120.7 \\
	 2.0   & -123.6 &   -91.4  & -90.2  & -89.6   &  -89.1 & -89.9  \\
	 3.0   & -105.5 &   -74.8  &        & -73.8   &  -73.7 & -73.7  \\ 
	 4.0   & -93.6  &   -64.0  &        & -63.6   &  -63.6 & -63.3  \\
	 5.0   & -85.0  &   -56.3  &-56.3   & -56.3   &  -56.3 & -55.8  \\
	 6.0   & -78.2  &   -50.5  &        & -50.7   &  -50.7 & -50.1  \\ \hline
	\end{tabular*}
\end{center}
\caption[]{Values of the correlation energy calculated from 
a variety of approximations and methods for the electron liquid, RPA, STLS \cite{70}, C-A,
VWN, EPX and the present interpolation method; C-A and VWN stand for
the original values Ceperley and Alder obtained from the Green's function
Monte Carlo method (GFMC) and the values from an interpolation formula
using their GFMC data, respectively; EPX stands for the effective potential
expansion method proposed by Takada \cite{35}.}
\end{table}

     The approximation method of replacing $\bar{g}({\bf r,r'})$ in the exact formula
for $E_{xc}$ in Eq. (\ref{eq:3.24}) by its analogue of the uniform electron liquid has
been proposed by a number of authors \cite{71,72,73,74,75,76}. This method has the merit of giving
the nonlocal effect of $E_{xc}$ in the sense that the resulting $v_{xc}({\bf r})$
implicitly depends on the electron density $n({\bf r'})$ at the neighboring
position $\bf r'$ of $\bf r$. As for the density parameter to be used in the function
$\bar{g}({\bf r-r'})$ of the electron liquid, we think that the uniform electron
density
in the interstitial region outside muffin-tin spheres for metals or the
averaged density of the system as a whole is rather better than the local
density. In this case it will be appropriate to approximate the functional
derivative $\delta\bar{g}({\bf r'-r''})/\delta n({\bf r})$
as $\delta({\bf r-(r'+r'')}/2)d\bar{g}({\bf r'-r''})/dn$
where $d\bar{g}({\bf r'-r''})/dn$ should be evaluated at the same density parameter as in
$\bar{g}({\bf r-r'})$.

\subsection{Some criticisms on the interpolation}
Finally, we offer some criticisms of the present interpolation method for
its possible development in the future. We have first interpolated between
long-range correlation in the RPA and short-range correlation in the
particle-particle ladder approximation in Sec. III C and then allowed
self-consistently for the exchange correction together with its feedback
effect on the direct interaction by iteration in Sec. III D. Once the local
field factor from short-range Coulomb repulsion $C({\bf q})$ introduced in Sec. III
C is given, there is no problem in the present method of evaluating the
exchange correction; the local field factor from exchange $G({\bf q})$ is uniquely
determined.

     The present theory depends on the accuracy of the starting
interpolation, i.e., the local field factor from short-range Coulomb
repulsion $C({\bf q})$. One of the authors (H.Y.) \cite{51} gave an analytic solution to the
ladder interaction $I({\bf p, p'; q})$ averaged over $\bf p$ and $\bf p'$ within the Fermi
sphere. We have employed this analytic solution for the determination of
the factor $C({\bf q})$ primarily for $q>p_F$ and made a smooth extrapolation to
smaller wavenumbers assuming an appropriate form quadratic in the limit
$q \to 0$.

     There remains some arbitrariness in
the choice of the local field factor $C({\bf q})$ around intermediate wavenumbers.
Those proper polarization functions of the direct type $\pi_b({\bf q},\omega)$ which
involve second- and
higher-order particle-particle ladder interactions make the 
predominant contribution to $S^{\uparrow \downarrow}({\bf q})$
and $\bar{S}^{\uparrow \downarrow}({\bf q})$ for intermediate and
large wavenumbers. It should, however, be pointed out that those proper
polarization functions of the same direct type which involve particle-hole
and hole-hole ladder interactions can make an appreciable contribution for
intermediate wavenumbers (see (3h) and (3j) in Fig. 1). By its strict definition, the 
local field factor from short-range Coulomb
repulsion $C({\bf q})$ should be determined from all the
proper polarization functions of the direct type $\pi_b({\bf q},\omega)$.
     
Since the first attempt by Hubbard \cite{14,15}, a number of interpolation methods
have been proposed with the local field correction to the RPA for the study of
the electron liquid in the region of metallic densities. The present theory
is the most sophisticated among these attempts. Here we have managed to
circumvent the treatment of muti-pair excitations \cite{22,23} by evaluating
the coupling-constant-averaged static structure factor $\bar{S}({\bf q})$ instead of
the static structure factor $S({\bf q})$. If one starts with 
$S({\bf q})$ in his interpolation, one cannot avoid treating multi-pair
excitations and thereafter has to perform the integral over the coupling
constant to enumerate correctly the many-body perturbation terms involving
particle-particle ladder interactions. Instead, we have started with the
exact ground state energy contribution from an infinite series of
particle-particle ladder diagrams in order to interpolate between the RPA
and the ladder approximation in the form of the coupling-constant-averaged
static structure factor $\bar{S}({\bf q})$.
\section{ORBITAL-DEPENDENT CORRELATION ENERGY FUNCTIONAL}
\subsection{Orbital-dependent exchange energy functional}
According to the optimized-effective-potential method (OEP) \cite{8} in DFT, the
ground state energy functional of many-electron systems can be written as
\begin{eqnarray}
E_g&=&\sum_{i=1}^{N}\int d{\bf r}\varphi^{*}_i({\bf r})\left(-\frac{\hbar^2}{2m}\nabla^2\right)\varphi_i({\bf r})\nonumber\\
&+&\int d{\bf r} v_{ext}({\bf r})n({\bf r})+\frac{1}{2}\int d{\bf r}\int d{\bf r'}\frac{n({\bf r})n({\bf r'})}{|{\bf r-r'}|}\nonumber\\
&-&\frac{1}{2}\sum_{i,j}\int d{\bf r}\int d{\bf r'}\frac{\varphi^{*}_i({\bf r})\varphi^{*}_j({\bf r'})\varphi_j({\bf r})\varphi_i({\bf r'})}{|{\bf r-r'}|}\nonumber\\
&+&E_c[\{\varphi_i\}].\label{eq:4.1}
\end{eqnarray}
The first term is the kinetic energy functional of the reference
noninteracting system with the same electron density $n({\bf r})$ as the real
system; this term is usually denoted by $T_s$. The second is the
electron-nucleus interaction energy functional. The third and
the fourth are the Hartree and the exchange energy functionals,
respectively. The exchange energy functional $E_x[\{\varphi_i\}]$ is represented as an explicit
functional of Kohn-Sham orbitals and hence $E_x[\{\varphi_i\}]$ is an
implicit functional of the electron density $n({\bf r})$; $i$ and $j$ have the same spin orientations. 
The sum of the third and fourth terms amounts to the electron-electron interaction energy functional
in the Hartree-Fock approximation. We assume that the
correlation energy functional is represented as an explicit functional of
Kohn-Sham orbitals. It is important to recognize that the correlation
energy functional $E_c[\{\varphi_i\}]$ defined in DFT amounts to the net correlational
lowering in the electron-electron interaction energy partially cancelled by
the correlational increase in the kinetic energy, $T_c$; $E_c=T_c+V_{ee}^c$ where $V_{ee}^c$ is the correlational
lowering in the electron-electron interaction energy.
   
  The Hohenberg-Kohn variational principle $\delta E_g[n({\bf r })]/\delta n({\bf r})$ $=0$ can be expressed
in terms of a self-consistent set of Kohn-Sham equations describing the
reference noninteracting system as
\begin{eqnarray}
&&\left\{-\frac{\hbar^2}{2m} \nabla^2 + v_{s}({\bf r})\right\}\varphi_i({\bf r})=\epsilon_i\varphi_i({\bf r})\label{eq:4.2}, \\ 
&&v_s({\bf r})=v_{ext}({\bf r})+v_H({\bf r})+v_x({\bf r})+v_c({\bf r}),\label{eq:4.3}\\
&&v_H({\bf r})=\int d{\bf r'}\frac{n({\bf r'})}{|{\bf r-r'}|},\nonumber\\
&&n({\bf r})=\sum_i^{occ.}|\varphi_i({\bf r})|^2,\nonumber
\end{eqnarray}
where $v_s({\bf r})$ denotes the total single electron potential; $v_{ext}({\bf r})$ is the
Coulomb potential from the nuclei and $v_H({\bf r})$ is the Hartree potential; $v_x({\bf r})$
and $v_c({\bf r})$ denote the exchange and the correlation potentials defined by
$v_x({\bf r})=\delta E_x[\{\varphi_i\}]/\delta n({\bf r})$
and $v_c({\bf r})=\delta E_c[\{\varphi_i\}]/\delta n({\bf r})$, respectively.

      In this subsection we mention only the exchange energy functional
$E_x[\{\varphi_i\}]$ and its functional derivative $v_x({\bf r})$ 
since it is a starting point
fundamental to the introduction of an orbital-dependent correlation energy
functional in DFT. An excellent review on the theoretical treatment of the
orbital-dependent exchange energy functional in DFT and its numerical
results for atoms, molecules and solids has recently been given by Grabo,
Kreibich, Kurth and Gross \cite{8}.
     
If one employs the orbital-dependent exchange energy functional, one
has to evaluate $v_x({\bf r})$ as follows:
\begin{eqnarray}
v_x({\bf r})&=&\delta E_x[\{\varphi_i\}]/\delta n({\bf r})\nonumber\\
&=&\sum_i \int d{\bf r'}d{\bf r''}\left\{\frac{\delta E_x[\{\varphi_i\}]}{\delta \varphi_i({\bf r'})}
\frac{\delta \varphi_i({\bf r'})}{\delta v_s({\bf r''})}+{\rm c.c}\right\}\frac{\delta v_s({\bf r''})}{\delta n({\bf r})},\nonumber\\
\label{eq:4.4}
\end{eqnarray}
where $\delta v_s({\bf r'})/\delta n({\bf r})$ is the inverse of the static density response function
of the reference noninteracting system, $\chi^{-1}({\bf r,r'})$. The functional
derivative $\delta \varphi_i({\bf r'})/\delta v_s({\bf r''})$ and the response function $\chi({\bf r,r'})$ can both be
expressed in terms of Kohn-Sham orbitals and Kohn-Sham energies. After a
few manipulations one can obtain the integral equation which determines the
exchange potential $v_x({\bf r})$.
\begin{eqnarray}
&&\sum_{i}^{occ.}\int d{\bf r'}\left\{\varphi^{\dagger}_i({\bf r'})G_{si}({\bf r',r})\varphi_i({\bf r})+{\rm c.c}\right\}v_x({\bf r'})\nonumber\\
&&=\sum_{i}^{occ.}\int d{\bf r'}\left\{\varphi^{\dagger}_i({\bf r'})\frac{1}{\varphi^{\dagger}_i({\bf r'})}\frac{\delta E_x[\{\varphi_i\}]}{\delta\varphi_i({\bf r'})}
G_{si}({\bf r',r})\varphi_i({\bf r})\right.\nonumber\\
&&~~~~~~~~~~~~~~~~~~~~~~~~~~~~~~~~~~~~~~~~~~~~~~~~~~~+{\rm c.c}\Bigg\},\label{eq:4.5}\\
&&G_{si}({\bf r,r'})=\sum_{i \neq j}^{\infty}\frac{\varphi_j({\bf r})\varphi_j^{\dagger}({\bf r'})}{\epsilon_i-\epsilon_j}\label{eq:4.6}.
\end{eqnarray}
It is straightfoward to generalize this integral equation to include the
correlation potential $v_c({\bf r})$ as well if an orbital-dependent correlation
energy functional $E_c[\{\varphi_i\}]$ is available.

      An alternative method can lead to the same integral equation for the
exchange-correlation potential $v_{xc}({\bf r}) (=v_x({\bf r})+v_c({\bf r}))$. Consider the ground
state energy functional $E_g$ in Eq.(\ref{eq:4.1}) under the condition that $\varphi_i({\bf r})$'s are
the solutions of Eq.(\ref{eq:4.2}) with an unknown one-electron potential
$v_s({\bf r})$. Then, $E_g$ can be considered to be a functional of $v_s({\bf r})$ and there
arises a variational problem of how to optimize $v_s({\bf r})$ in the minimization
of $E_g$ , i.e., $\delta E_g[v_s({\bf r})]/\delta v_s({\bf r})=0$. The potential $v_s({\bf r})$ thus calculated is named
the optimized-effective-potential (OEP). From this optimization one can
also derive the same integral equation that we have just mentioned above.
Hence it is termed the OEP integral equation. In fact, the optimization
$\delta E_g[v_s({\bf r})]/\delta v_s({\bf r})=0$ is equivalent to the Hohenberg-Kohn variational principle
$\delta E_g[n({\bf r})]/\delta n({\bf r})=0$ \cite{77,78}.
We summarize the conclusions obtained from the so-called exchange-only OEP method in the following.

(1) The exchange-only OEP method gives the exact exchange potential
$v_x({\bf r})$ that completely cancels the self-interaction terms involved in the
Hartree potential.

(2) The derivative discontinuity 
in $E_x[\{\varphi_i\}]$, or the discontinuity of
$v_x({\bf r})$ can be correctly described.

(3) The Hartree-Fock ground state energy in the framework of DFT is by
only a slight amount higher than the usual Hartree-Fock ground state
energy. This is because in DFT the additional condition of a local form is
imposed on the exchange potential in the minimization of the same ground
state energy functional.

(4) Instead, the electron density $n({\bf r})$ calculated from DFT might be by
only a slight amount better than the one calculated from the usual
Hartree-Fock theory owing to the variational principle. 

However, it is important to recognize that the exchange-only OEP integral equation is
identical to the first iterative form of the exchange-only Sham-Schluter
equation (see Eq.(\ref{eq:5.4})). The Sham-Schluter equation implies that the
standard many-body theory based on the one-electron Green's function
formalism and DFT both lead to the same exact electron density $n({\bf r})$ of
many-electron systems. This requirement of course should be satisfied only
when both of the two many-body theories are exact. If applied to the
Hartree-Fock stage of approximation, this equation amounts to forcibly
requiring that the Hartree-Fock approximation in the framework of DFT lead
to the same electron density $n({\bf r})$ as the usual Hartree-Fock theory; it
cannot be fulfilled. From the first iterative form of the exchange-only
Sham-Schluter equation it is evident that the difference in
the electron density $n({\bf r})$ between the two methods will be only a slight
amount just like the difference in the ground state energy.

(5) Li, Krieger and Iafrate \cite{79} have applied the exchange-only
spin-unrestricted OEP method to atoms (Z=3$\sim$56(Li$\sim$Ba), $79$ and 
80 (Au and Hg)) to make a comparison with the spin-unrestricted Hartree-Fock method
(SUHF). Their calculations are very accurate; the virial theorem and the
related identity
$(V_{ee}^x+ \int d{\bf r}n({\bf r}){\bf r}\cdot {\rm grad} v_x({\bf r})=0)$ 
for the exchange energy functional $E_x(=V_{ee}^x)$ and the exchange potential 
$v_x({\bf r})$ due to Levy and Perdew \cite{80} are satisfied to four or five significant figures.

The ground state energy calculated from  the exchange-only OEP
method is at most by 40ppm higher than that from the  Hartree-Fock method
and the differnce decreases with increasing Z. They report that the
electron density $n({\bf r})$ calculated from the exchange-only spin-unrestricted
OEP method is close to the SUHF value. Interestingly, they observe a few
reversals of the order of the state yielding the highest occupied
eigenvalue between the $n{\rm s}$ and $(n-${\sl 1}){\rm d} states in the transition series of
atoms. The Kohn-Sham exchange potential
$v_x({\bf r})$ always raises the energy levels of the inner shell states compared
with those of the SUHF; the same trend can also be seen from a comparison
between the Hartree-Fock energy levels of occupied states of the electron
liquid and the corresponding Kohn-Sham energy levels. The reversals occur only when
the $n{\rm s}$ and $(n-${\sl 1}){\rm d} energy levels are very close in the SUHF. Along the
spirit of the Slater approximation to the exchange potential, Krieger et al
(KLI) \cite{81,82,83} have proposed a simple approximate solution to the OEP integral
equation in which the complexity of the original OEP integral equation is
significantly reduced but many of its essential properties are kept
unchanged.

(6) Buijse and Baerends \cite{84} write that the Hartree-Fock theory
significantly distorts the electron density distribution $n({\bf r})$ of atoms to
the extent that often the largest absolute errors in the total Hartree-Fock energy
occur for the electron-nucleus interaction energy and the kinetic energy
rather than for the electron-electron interaction energy and that the
Slater approximation for the exchange potential gives rather better results
for the electron density $n({\bf r})$ of atoms. However, the largest relative
error, i.e., the largest ratio of each error to the corresponding Hartree-Fock value
probably will occur for the electron-electron interaction energy.

(7) Hartree-Fock calculations make higher {\sl 4}s orbitals occupy prior to
lower {\sl 3}d orbitals in {\sl 3}d transition atoms. This reversal of the occupied
states has been supposed to be ascribed to the variational principle
$\delta E_g^{HF}[\{\varphi_i\}]/\delta \varphi_i=0$. 
The reversal has been interpreted to be a consequence of
the strategy that the total energy will not be raised due to the
electron-electron interaction energy in $E_g^{HF}[\{\varphi_i\}]$. On the other hand, no
reversal of occupied states should occur if one accurately solves the
exchange-only OEP integral equation in DFT. This is because Kohn-Sham
equations by definition are intended for the reference non-interacting
system. This problem may be closely related to the central field approximation
conventionally applied to atoms. However, no systematic method has yet been
developed to solve the OEP integral equation appropriately for a
nonspherically symmetric system. Strictly, the degeneracy of {\sl 3}d or {\sl 4}d
orbitals should be lifted for lack of spherical symmetry and $n{\rm s}$ and $(n-${\sl 1})d
orbitals should be mixed as a result.

(8) As for the energy gap of semiconductors, the exchange-only OEP
method gives a good result \cite{85,86}.

(9) Strictly, the Hartree-Fock theory and its density-functional version by the x-only
OEP method generally give different results of the ground state energy $E_g$, the electron 
density $n({\bf r})$ and the ionization potential ( or the chemical potential ); for the 
uniform electron liquid the two give the same result of these quantities.
\subsection{Correlation-induced reconstruction of Kohn-Sham orbitals}
In this subsection we propose an explicitly orbital-dependent correlation
energy functional beyond the Hartree-Fock approximation for the study 
of the electronic structure in atoms, molecules and solids in DFT. It
is to be employed in combination with the orbital-dependent exchange energy
functional.

     In Sec.III we have defined a new expression for the correlation energy
of the electron liquid. It is a modification of second-order perturbation
terms, in which one of the two Coulomb interactions in each term is
replaced by an effective interaction containing information about long-,
intermediate-, and short-range correlations. Similarly, we propose here a
second-order perturbation like expression \cite{87} which is constructed from
Kohn-Sham orbitals, Kohn-Sham energies, and the effective interaction
borrowed from the electron liquid.
\begin{widetext}
\begin{eqnarray}
&&E_c[\{\varphi_i\}]=\frac{1}{2}\sum_{i,j}^{occ.}\sum_{a,b}^{unocc.}\left\{
\frac{\langle\varphi_i\varphi_j|v({r}_{12})|\varphi_a\varphi_b\rangle
\langle\varphi_a\varphi_b|v_{eff}({r}_{12}) |\varphi_i\varphi_j\rangle}
{\epsilon_i+\epsilon_j-\epsilon_a-\epsilon_b}
-\frac{\langle\varphi_i\varphi_j|v({r}_{12}) |\varphi_a\varphi_b\rangle
\langle\varphi_a\varphi_b|v_{eff}({r}_{12}) |\varphi_j\varphi_i\rangle}
{\epsilon_i+\epsilon_j-\epsilon_a-\epsilon_b}\right\}\nonumber,\\\label{eq:4.7}\\
&&\langle\varphi_i\varphi_j|v(r_{12})|\varphi_a\varphi_b \rangle = \int d{\bf r}_1d{\bf r}_2
\varphi_i^*({\bf r}_1)\varphi_j^*({\bf r}_2)v(r_{12})\varphi_a({\bf r}_1)\varphi_b({\bf r}_2),\label{eq:4.8}\\
&&\langle\varphi_i\varphi_j|v_{eff}(r_{12})|\varphi_a\varphi_b \rangle = \int d{\bf r}_1d{\bf r}_2
\varphi_i^*({\bf r}_1)\varphi_j^*({\bf r}_2)v_{eff}(r_{12})\varphi_a({\bf r}_1)\varphi_b({\bf r}_2),\label{eq:4.9}
\end{eqnarray}
\end{widetext}
where $i,j$ and $a,b$ represent occupied and unoccupied Kohn-Sham states,
respectively and the states $i,j$ in the second-order exchange like term have
the same spin orientations; $v(r_{12})$ is the Coulomb interaction for $r_{12}= |{\bf r_1-r_2}|$
 The effective interaction $v_{eff}(r_{12})$ in Eq.(\ref{eq:4.7}) is the real-space
Fourier transform of the interaction $v_{eff}({\bf q})$ we have evaluated for the
electron liquid in Sec.III. How to choose in practical calculations the
density parameter for the effective interaction we borrow from the
electron liquid will be discussed in detail in Sec. IV C.

     Here we note that a fully orbital-dependent correlation energy
functional, or equivalently an orbital-dependent effective interaction in
the present scheme can in principle be obtained if one follows the same
procedure as in the electron liquid with Kohn-Sham orbitals and Kohn-Sham
energies in place of planewaves and free electron energies.

     From a comparison with a general expression for the
exchange-correlation energy functional using the coupling-constant-averaged
pair correlation function $\bar{g}({\bf r,r'})$ (see Eq.(\ref{eq:3.24})) it may be naturally
expected that the present correlation energy functional has also the effect
to reduce the Hartree-Fock electron-electron interaction energy in its
contributions from short distances.
     Let us characterize the present orbital-dependent correlation energy
functional.

     (1) The exact many-electron wavefunction fulfills the requirement of
antisymmetry. In accordance with this antisymmetry, the electron-electron
interaction energy functional in DFT should be constructed from direct and
exchange pairs of terms. The present orbital-dependent correlation energy
functional as well as a direct and exchange pair of electron-electron
interaction energy functionals in the Hartree-Fock approximation reflects
the requirement of antisymmetry.

     (2) In the limit of uniform density the present theory is reduced to
the local density approximation (LDA) as it should be. The self-interaction
terms involved in the Hartree energy are exactly cancelled by the
orbital-dependent exchange energy
functional. Furthermore, higher-order self-interaction terms are, from the
beginning, eliminated from the present $E_c[\{\varphi_i\}]$ since it consists of a
direct and exchange pair of terms. Therefore the present theory is
completely free from self-interaction errors. This is indispensable for the treatment of exchange and
correlation between electrons with localized orbitals. The LDA estimates
the location of the occupied {\sl 3}d-bands in copper and zinc too highly
because of self-interaction errors. The overestimation as large as about
$18\%$ in the neck cross section of the Fermi surface of copper in the [111]
direction, which is considered to be ascribed to the too high {\sl 3}d-bands in
the LDA, will be for the most part improved by removing self-interaction
errors \cite{88}.

     (3) One of the remarkable merits of the present theory is that the
physical meaning of correlation is clarified by the present representation
of the functional $E_c[\{\varphi_i\}]$. According to the virial theorem, the
correlation energy of atoms, molecules and solids under no constraint of any 
external force amounts to a half of the
lowering in the total potential energy since it is by half cancelled by the
correlational increase in the kinetic energy. In this sense the partitioning
of the ground state energy functional in DFT is helpful in understanding
the meaning of the correlation energy since it is an exact many-body theory
assuming the self-consistent one-electron formalism. The total potential
energy consists of three terms. The first is the electron-electron interaction
energy and the second the electron-nucleus interaction energy. The third is
the nucleus-nucleus interaction energy. Usually, the first and second potential
energies are both lowered at the cost of increasing the corresponding
two kinetic energies, $T_c$ and $T_s$. Here we assume the Born-Oppenheimer
approximation.

     It is important to notice that correlation usually changes the
magnitude of the nucleus-nucleus interaction energy from its Hartree-Fock
value. The Hartree-Fock
approximation regularly underestimates the equilibrium length between the
nuclei in molecules. The LDA that borrows knowledge of the exchange-correlation
energy from the uniform electron liquid has a tendency to underestimate the
Wigner-Seitz radius of metals or the lattice constant of solids. Probably,
these shortcomings will be closely related to the fact that both
approximations overestimate the screening of the nuclei at short distances.

     By definition, the functional $E_c[\{\varphi_i\}]$ in DFT represents the net 
lowering of the electron-electron interaction energy partially cancelled by the
correlational increase in the
kinetic energy $T_c$. From a comparison with the Hartree-Fock total energy
functional it is evident that the direct term in Eq.(\ref{eq:4.7}) has the effect to
reduce the
Hartree energy and its exchange counterpart has the effect to reduce in
magnitude the exchange energy in a well-balanced way through the same
effective interaction $v_{eff}({\bf r})$ that contains accurate information about
long-, intermediate-, and short-range correlations beyond second-order
perturbation theory.

     Furthermore, it is important to recognize that the presence of
$E_c[\{\varphi_i\}]$ in the total energy functional not only has the effect to reduce
the net electron-electron interaction energy in its contributions from
short distances but also causes a change in the electron density $n({\bf r})$ such
that the total energy is further stabilized by a lowering in the only
negative energy, i.e., the electron-nucleus interaction energy
though it inevitably increases the kinetic energy of the reference
non-interacting system $T_s$. The partitioning of the ground state energy
functional in DFT has the merit of connecting the net lowering in the electron-electron
interaction energy (in its contributions from short distances) with the
further lowering in the electron-nucleus interaction energy through
Kohn-Sham equations that minimize the total  energy functional.

    (4) The present theory is expected to give an accurate evaluation of
the correlation energy even of the most localized electrons such as atomic
electrons in which the LDA makes the most serious errors because of its
failure to cancel self-interaction terms and its overestimation of the
correlation energy. Then we may expect that the present $E_c[\{\varphi_i\}]$ is valid
for tightly binding electrons as well as for nearly free electrons. These
two different types of electrons are both participating in the cohesion of
transition metals. An accurate evaluation of the ground state energy of
atoms, molecules and solids may be generally expected from the present
theory if the density parameter for the effective interaction
borrowed from the electron liquid is suitably chosen. The optimization of
the density parameter will be discussed in detail in Sec. IV C.

     (5) One of the greatest merits of DFT lies in the fact that one can
describe all correlation effects on one-electron states with the local
correlation potential ${v_c({\bf r})}$ in the framework of self-consistent
one-electron theory, without being troubled with the treatment of 
the non-local, energy-dependent, and complex self-energy operator involving 
many-body background excitations. This merit is
acquired only when accurate information about the potential ${v_c({\bf r})}$ is
available. The total single electron potential ${v_s({\bf r})}$ in Kohn-Sham equations
determines the electron density $n({\bf r})$ of the system. Its accuracy depends
entirely on the exchange-correlation potential $v_{xc}({\bf r})$. The exchange
potential $v_x({\bf r})$ can be exactly treated by the exchange-only OEP method.
Consider what effect the correlation potential $v_c({\bf r})$ has on the electron
density $n({\bf r})$ of the system.

     The correlated motion of valence electrons in solids, though it inevitably increases the
kinetic energy, more lowers the electron-electron interaction energy in its
contributions from short distances. The simultaneous change in the electron
density $n({\bf r})$, i.e., the shrinkage of the electron distribution around the
nuclei arising from less screening of the nuclei due to short-range
correlation, further lowers the only negative electron-nucleus interaction
energy at the cost of increasing the kinetic energy of the reference
non-interacting system $T_s$. Then the lowest possible ground state energy is
realized such that the nucleus-nucleus interaction energy is lowered by
somewhat increasing the lattice constant as a counteraction of the
shrinkage of the electron distribution around the nuclei and the increase
in $T_s$ is to some extent relaxed accordingly.

     Thus, correlation in real solids is necessarily accompanied
by a change in the electron density $n({\bf r})$ in contrast with the case of the
uniform electron liquid. It may be generally expected that correlation
induces inhomogeneity in the electron density distribution $n({\bf r})$ of the
system against the Hartree-Fock approximation in DFT.

     In the neighborhood of individual nuclei of the system, the total
single electron potential $v_s({\bf r})$ in Kohn-Sham equations can be written in
terms of the coupling-constant-averaged pair correlation function
$\bar{g}({\bf r,r'})$ as
\begin{eqnarray}
v_s({\bf r})&=&-\frac{Ze^2}{ |{\bf r}| }+\int d{\bf r}'\frac{n({\bf r}')}{|{\bf r-r'}|}\nonumber\\
	    &+&\int d{\bf r'}\frac{n({\bf r'})}{ |{\bf r-r'}| }\{\bar{g}({\bf r,r'})-1\}\nonumber\\
            &+&\frac{1}{2}\int d{\bf r'} d{\bf r''}\frac{n({\bf r'})n({\bf r''})}{|{\bf r'-r''}|}
\frac{\delta \bar{g}({\bf r',r''})}{\delta n({\bf r})}.\label{eq:4.10}
\end{eqnarray}
The screening of individual nuclei at short-, intermediate-, and long-distances 
can be properly described by the second and third terms; for
isolated systems like atoms these two terms give the correct asymptotic
form $v_s({\bf r})=-e^2/|{\bf r}|$ for large $|{\bf r}|$ since the following sum rule holds:
\begin{eqnarray}
\int d{\bf r'}n({\bf r'})\{\bar{g}({\bf r,r'})-1\}=-1.\label{eq:4.11}
\end{eqnarray}
The last term in Eq.(\ref{eq:4.10}) probably will not give a significant ${\bf r}$-dependence
for any $|{\bf r}|$ and plays a background-like role in $v_s({\bf r})$ since the integrand 
including the functional derivative
$\delta \bar{g}({\bf r',r''})/\delta n(\bf r)$ is integrated over $\bf r'$ and $\bf r''$. Consider the physical
role of $\bar{g}({\bf r,r'})$ in the sum of the second and third terms when the
distance $|{\bf r-r'}|$ is decreased. Since the inequality that $0< \bar{g}({\bf r,r'})< 1/2$ holds
for $|{\bf r-r'}|$ less than the average inter-electron separation, it is
evident that the Hartree or Hartree-Fock screening of individual nuclei is significantly
reduced at
short distances due to the presence of $\bar{g}({\bf r,r'})$. In other words,
short-range correlation yields less screening of the attractive nuclear
potential at short distances. It is this less screening of the nuclear
potential that shrinks the electron distribution in the vicinity of the nuclei to
enhance the electron density $n({\bf r})$ at the positions of individual nuclei.

If $\bf r$ and $\bf r'$ are both in the
interstitial region
outside muffin-tin spheres of metals, the function $\bar{g}({\bf r,r'})$ is very well
approximated by the corresponding function of the uniform electron liquid
we have calculated in Sec. III E and hence causes no change in the electron
density. On the other hand, if $\bf r$ and $\bf r'$ are both inside a muffin-tin
sphere, the probability of $\bar{g}({\bf r,r'})$ is concerned with not only valence electrons but also
core electrons. The magnitude of $\bar{g}({\bf r,r'})$ at short separations within the
muffin-tin sphere is less than $1/2$ due to short-range correlation, though
it may be somewhat greater than in the interstitial region. Short-range correlation 
in this essentially nonuniform density region reduces the Hartree or Hartree-Fock screening
of the nuclei and hence causes the shrinkage of the electron density
distribution around the nuclei to lower the electron-nucleus interaction
energy more than its Hartree-Fock value.

(6) Most importantly, the present orbital-dependent correlation energy
functional $E_c[\{\varphi_i\}]$ hybridizes Hartree-Fock Kohn-Sham orbitals in order to
produce self-consistently reconstructed Kohn-Sham orbitals through its
functional derivative $\delta E_c[\{\varphi_i\}]/\delta n({\bf r})$, i.e., the correlation potential
$v_c({\bf r})$. Naturally, this correlation-induced reconstruction of Kohn-Sham
orbitals gives a further stabilization of the ground state energy
as well as a redistribution of the electron density $n({\bf r})$ of the system.
The present correlation potential $v_c({\bf r})$ is expected to produce less screening of the
attractive nuclear potential at short distances to give a shrinkage of $n({\bf r})$ around
individual nuclei, thereby leading to a lowering in the electron-nucleus interaction energy.

The Hartree-Fock approximation, in its original version or its DFT version, 
overestimates the effect of the electron-electron interaction on the screening of the nuclei at short
distances as well as the magnitude of the electron-electron interaction
energy. Variation of the expectation value of the total Hamiltonian within a single Slater determinant
overestimates the influence of the interaction on occupied orbitals in the form of self-consistent 
Coulomb and (nonlocal or local) exchange potentials. The Hartree-Fock orbitals are then 
spread too far away from the nuclei and hence the kinetic and the
electron-nucleus interaction energies in this approximation are both underestimated in magnitude. 

From these peculiarities it is evident that this approximation tends to underestimate nucleus-nucleus 
separations in order to get a sufficient lowering in the only negative electron-nucleus interaction energy 
for the stabilization of molecules.
There are some cases such as ${\rm F}_2$, in which  the Hartree-Fock approximation is unable to 
stabilize molecules and the redistribution of the
electron density due to correlation is indispensable to the formation of
molecules.

The LDA is exact in the limit of uniform density but, when applied to
nonuniform electronic systems, it cannot describe properly such
exchange-correlation effects as are peculiar to nonuniform systems. The
most important of these effects is correlation-induced less screening of
the nuclei at short distances. The LDA tends to underestimate the lattice
constant of solids for lack of this effect and furthermore tends to
overestimate the magnitude of the cohesive energy because of a rather too
uniform spatial variation of the exchange-correlation potential $v_{xc}({\bf r})$. Notice 
that the LDA considerably smooths out the spatial variation of $v_{xc}({\bf r})$ 
since it borrows the knowledge of $\mu_{xc}(n({\bf r}))$ from the
uniform electron liquid; $\mu_{xc}( n({\bf r}) )$ is the exchange-correlation 
contribution of the chemical potential of the uniform electron liquid appropriate for the
electron density $n({\bf r})$. 

(7) On the basis of the virial theorem and his own partitioning of the total energy,
Ruedenberg has gained a profound insight into the binding nature of
molecules \cite{89,90}. He has emphasized that the binding of molecules or the cohesion
of solids (including metals) is essentially stabilized by shrinking the
electron density distribution in the immediate vicinity of the nuclei to enhance the electron
density $n({\bf r})$ at the positions of indivisual nuclei, compared with the case of
isolated atoms.

According to Moruzzi, Janak and Williams \cite{7}, on the other hand, the
electron density in simple metals is reversely reduced in the immediate
vicinity of the nuclei. This is because the LDA fails to give the proper
description of the correlation-induced less screening of individual nuclei
at short distances. In the LDA the
Hartree screening of the nuclear potential is simply shifted by an amount
of $\mu_{xc}( n({\bf r}) )$. The density dependence of $\mu_{xc}( n({\bf r}) )$ 
appropriate for the electron liquid is absolutely insufficient for describing the 
correlation-induced less screening of nuclei. It is beyond the
scope of the LDA since the electron density $n({\bf r})$ in the vicinity of the
nuclei in metals is far from being slowly varying.

The present theory is promising for the realization of the enhancement
of the electron density in the vicinity of the nuclei as a consequence of
less screening of the nuclear potential. This is because the present $E_c[\{\varphi_i\}]$ involves
all unoccupied excited states that are needed to 
redistribute the electron density $n({\bf r})$ in this region; the resulting $v_c({\bf r})$
is expected to have much larger negative values than the LDA correlation potential
in this region.

Kohn-Sham one-electron states have to be reconstructed while atoms
condense into solids. We think that the reconstruction of Kohn-Sham
one-electron states associated with the cohesion of solids can also be well
described by the present correlation energy functional provided the density
parameter in the effective potential is appropriately chosen.

In connection with the considerations above we mention a noteworthy fact.
According to Boyd and his coworkers \cite{91,92,93}, Hund's multiplicity rule can be
ascribed to a gain in the electron-nucleus potential energy caused by
correlation-induced less screening of the nuclei in contrast with the
traditionally accepted interpretation due to the exchange energy. On the
contrary, the electron-electron interaction energy is grater in a high
spin state for all neutral atoms and molecules calculated with requisite
accuracy. They have arrived at this conclusion from accurate configuration
interaction calculations of low-lying excitation energies of light atoms
and molecules, in which the virial theorem is of course satisfied.

(8) There is a strong possibility that the electronic structure of the
so-called strongly correlated electron systems may be interpreted in terms
of the correlation-induced reconstruction of Kohn-Sham orbitals. The energy-band structure of
solids calculated in the LDA is approximately equivalent to the
band-structure that is
calculated in the Hartree approximation and shifted to the low energy side
by an amount of $\mu_{xc}$ of the uniform electron liquid with the averaged
electron density of solids. In marked contrast to the LDA, the present
orbital-dependent correlation energy functional hybridizes Hartree-Fock
Kohn-Sham one-electron states with different orbital characters,
 particularly in the neighborhood of the Fermi level or
the energy gap, in order to produce self-consistently reconstructed
Kohn-Sham one-electron states. The occurrence of such reconstruction of
one-electron states is peculiar to non-uniform real systems, quite absent
in the uniform electron liquid which consists of electrons with the same character.

Consider the energy denominator in the present $E_c[\{\varphi_i\}]$. Evidently,
the functional $E_c[\{\varphi_i\}]$ can gain its magnitude if the density of states is
sharply enhanced in the immediate neighborhood of the Fermi level under the
influence of self-consistently reconstructed Kohn-Sham energies. Such an
enhancement in the density of states should be accompanied by a considerable change in
the electron density  $n({\bf r})$ particularly for strongly anisotropic systems.

Our scenario of heavy fermions is as follows: As a consequence of
strong hybridization between conduction electrons and f electrons, the occupied
and unoccupied states will probably be reconstructed such that they enhance
the density of states in the immediate neighborhood of the Fermi level as
sharply as possible in order to gain the magnitude of the functional
$E_c[\{\varphi_i\}]$, which is balanced against a simultaneous increase in the kinetic
energy of the reference non-interacting system.

For the study of the electronic structure of heavy fermion systems it
is of course indispensable to adopt the fully relativistic one-electron
Dirac equations \cite{94} in the framework of DFT.

(9) Almbladh and von Barth \cite{95} have studied the relations between
Kohn-Sham eigenvalues and exact excitation energies. From exact asymptotic
results for the spin density far away from a finite system, they have shown
that the uppermost Kohn-Sham eigenvalue equals the ionization potential,
not only for extended systems like solids, but also for finite systems like
atoms.

From available accurate wavefunctions for a number of light atoms (the
He isoelectronic series, Li and Be), Almbladh and Pedroza \cite{96} have constructed
$E_{xc}$, $v_{xc}({\bf r})$ and $v_c({\bf r})$ to evaluate the Kohn-Sham eigenvalues. They have then
found that the uppermost (or highest occupied) Kohn-Sham eigenvalue
reproduces the exact ionization potential and that the eigenvalues for
deeper shells lie above the corresponding exact excitation energies, while
the Hartree-Fock eigenvalues lie below.

The correlation potential $v_c({\bf r})$ they have constructed for He, ${\rm Li}^+$ and
${\rm Be}^{++}$ takes large negative values in the immediate vicinity of the nucleus
reflecting that the exact electron density is slightly higher than the
Hartree-Fock value in this region and on the other hand becomes positive at
large $r$ from the nucleus, forms a broad peak and tends to zero with
increasing $r$. This is quite different from the potential $v_c({\bf r})$ in the LDA
since it is always negative and increases monotonically with increasing $r$.

Generally, the Koopmans theorem gives a fairly good evaluation of the
ionization potential for atoms and molecules, though Hartree-Fock
eigenvalues take no account of both correlation and relaxation \cite{97}. This is
because a great amount of cancellation occurs between the extra correlation
energy of the $N$-body ground state relative to the $(N-1)$-body excited state
and the relaxation energy (which is also a kind of correlation energy) of
the $(N-1)$-body excited state produced by removing an electron from the
Hartree-Fock N-body ground state.

This cancellation is not complete, however, and the Koopmans theorem
usually overestimates the ionization potential, or equivalently the
Hartree-Fock theory tends to evaluate the highest occupied eigenvalue
(negative) too low. This implies that the relaxation energy is usually
somewhat larger than the extra correlation energy in magnitude; for much
localized orbitals like the He atom it is so large that the Koopmans
theorem cannot be applied. Thus, the correlation correction to the
ionization potential is usually negative for molecules; the Hartree-Fock Kohn-Sham 
equations with the optimized local exchange potential will lead to the same conclusion.
The presence of nuclei in atoms and molecules is
responsible for the occurrence of relaxation. If an electron is removed, the
remaining $N-1$ electrons have to readjust themselves to the attractive
nuclei.

It is very reasonable that the correlation potential $v_c({\bf r})$ for He, ${\rm Li}^+$
and ${\rm Be}^{++}$  takes positive values at large distances from the nucleus if one
considers that the expectation value of $v_c({\bf r})$ with respect to the highest
occupied state has to give a correction with the positive
sign.

On the other hand, the solutions of Hartree-Fock or Hartree-Fock
Kohn-Sham equations for the uniform electron liquid are plane waves and
hence there occur no
relaxation of orbitals. Then, only the extra correlation energy is missing
in this case. The correlation contribution to the chemical potential, $\mu_c$
for the uniform electron liquid is thus negative. 　　　　　　　　        　

Averill and Painter \cite{98}, and Levy and Perdew \cite{80} have given an expression for
the virial theorem in terms of density-functional theory. Their expression
for the virial theorem at the equilibrium positions of the nuclei is
splitted into two parts, i.e., the reference one-electron (non-interacting)
and the correlational parts.
\begin{eqnarray}
2T_c+V_{ee}^c+\int d{\bf r}n({\bf r}){\bf r}\cdot{\rm grad}v_c({\bf r})=0.\label{eq:4.12}
\end{eqnarray}
This is the correlational part of the virial theorem that relates the
correlational increase in the kinetic energy $T_c$ , the correlational
lowering in the electron-electron repulsion energy $V_{ee}^c$ , and the
correlation potential $v_c({\bf r})$; $T_c$ is usually incorporated into the
exchange-correlation energy functional $E_{xc} (E_{xc}=T_c+V_{ee}^c)$.
This equation is helpful to discuss the general shape of $v_c({\bf r})$.

Let us first consider the quantity $2T_c+V_{ee}^c$ in Eq.(\ref{eq:4.12}). It vanishes to
second order in $e^2$ for atoms and molecules. It also vanishes for the
electron liquid in the high density region where the correlation energy is
given by the leading logarithmic and the subsequent constant terms. Consider
the contribution to $2T_c+V_{ee}^c$ from very large momentum transfer interactions
in the electron liquid at metallic densities, or very short-range parts of
the Coulomb interaction. It also vanishes since the relation $2T_c+V_{ee}^c=0$
holds for this case \cite{54}. The reason for this cancellation is that the
particle-particle ladder interaction $I({\bf p,p'; q})$, or multiple scattering
between two particle states with opposite spins makes a contribution to
both $T_c$ and $V_{ee}^c$ in a way analogous to second-order perturbation
theory. This type of cancellation between $2T_c$ and $V_{ee}^c$, associated with
short-range correlation, applies also to atoms, molecules and solids.

From these considerations, $2T_c+V_{ee}^c$ is a rather small but negative
quantity. It is negative for the electron liquid at metallic densities. Second-order
perturbation theory for atoms and molecules, by which $2T_c+V_{ee}^c$ vanishes,
tends to overestimate the magnitude of the correlation energy and hence the
total higher-order contribution to $2T_c+V_{ee}^c$ is expected to be negative. We
may then conclude from Eq.(\ref{eq:4.12}) that the integral  
$\int d{\bf r}n({\bf r}){\bf r}\cdot {\rm grad}v_c({\bf r})$ is a small 
positive quantity for atoms, molecules and solids (including metals).

Correlation changes the electron density $n({\bf r})$ in atoms,
molecules and solids. The correlation potential $v_c({\bf r})$ is responsible for this change
in $n({\bf r})$. Short-range correlation rectifies the overestimated screening of
the nuclei in the Hartree- or Hartree-Fock approximation and enhances the
electron density $n({\bf r})$ in the immediate vicinity of the nuclei as a result.
Correspondingly, the potential $v_c({\bf r})$ takes large negative values in this
region. As the distance from the nucleus r increases, the value of $v_c({\bf r})$
increases with a steep positive slope. In order to conserve total
electronic charge, the potential $v_c({\bf r})$ for an atom has to become positive
for larger $r$ from the nucleus, forms a broad peak and tends to zero with
increasing $r$. The correlation potential $v_c({\bf r})$ that Almbladh and Pedroza
have constructed from a few accurate many-electron wavefunctions actually
shows such a nonmonotonic behavior.

Similarly, the correlation potential $v_c({\bf r})$ for metals will probably
take large negative values in the immediate vicinity of each nucleus and steeply 
increase with the distance from the nucleus. It is expected to exceed 
$\mu_c(r_s^*)$ (instead of zero) within the muffin-tin sphere as the distance from the nucleus $r$ is increased,
form a broad peak and tend to $\mu_c(r_s^*)$ as $r$ approaches the muffin-tin sphere radius;
$\mu_c(r_s^*)$ is the correlation contribution of the chemical potential of the
uniform electron liquid with the density parameter $r_s^*$ appropriate for the
interstitial region of metals. It should be noted that the potential $v_c({\bf r})$
in the immediate vicinity of nuclei probably will take larger negative values
for metals than for isolated atoms, reflecting that the electron density in this region
should be enhanced under the influence of cohesion.

The tendency of the LDA to overestimate the cohesive energy of metals
in magnitude and to underestimate the lattice constant of metals provably
will be ascribed to the failure of its correlation energy functional to
give the above nonmonotonic behavior of the correlation potential $v_c({\bf r})$ as
well as to the failure of its exchange energy functional to cancel spurious
self-interaction terms.

(10) The correlation energy defined in DFT, $E_c^{\rm DFT}$ is in magnitude
slightly larger than the usual correlation energy, $E_c$ defined by the
differnce between the usual Hartree-Fock and the exact ground state energies (see
IV A. (3)). According to the virial theorem,
$E_c^{\rm DFT}=-(T_s-T_s^{\rm HF})-T_c=\frac{1}{2}(V_{es}+V_{ee}^x+V_{ee}^c)-\frac{1}{2}(V_{es}+V_{ee}^x)^{\rm DFHF}$ 
where $V_{es}$ is the total classical electrostatic energy that consists of the nucleus-electron,
the Hartree, and the nucleus-nucleus interaction
energies; $V_{ee}^x$ and $V_{ee}^c$ are the exchange and correlation contributions to
the electron-electron interaction energy, respectively. The inequalities
that $T_s>T^{HF}>T_s^{HF}$ hold since two different virial theorems with the same
formal expression are satisfied for the usual Hartree-Fock approximation
and its density-functional version (see Eq.\ref{eq:4.13}). The quantity $T^{HF}$ stands for the kinetic
energy of the usual Hartree-Fock theory.

We give an explanation of the correlation energy defined in DFT, $E_c^{DFT}$.
The first term, $-(T_s-T_s^{HF})$ is the negative of an increase in the kinetic
energy of the reference non-interacting system, which is primarily associated with a
correlation-induced enhancement in the electron density in the immediate
vicinity of nuclei, giving a correlational lowering in the
nucleus-electron and nucleus-nucleus interaction energies. The second term,
$-T_c$ , on the other hand, is the negative of an increase in the kinetic
energy arising from the correlated motion of many electrons, which is
primarily associated with a correlational lowering in the electron-electron
interaction energy.

\subsection{Density parameter appropriate for the effective interaction}
In Sec. III B we have mentioned that a fully orbital-dependent correlation
energy functional, or an orbital-dependent effective interaction can in
principle be obtained if one makes a similar interpolation between long-
and short-range correlations involving exchange effects with Kohn-Sham
orbitals and Kohn-Sham energies in place of planewaves and free-electron
energies for the electron liquid. At present, such a sophisticated interpolation is
exceedingly difficult to perform. Instead, we borrow the knowledge
of the effective interaction from the electron liquid. In this case there
arises a problem of how to choose the density parameter for the effective
interaction. We must then resolve this problem to make the present theory
applicable to practical calculations. For this purpose we adopt the
following method.

Moruzzi, Janak and Williams \cite{7} were the first who successfully applied DFT to
energy-band calculations of metals with the atomic number less than
50 using the LDA. One of the most important conclusions they
have obtained from their systematic study is that the compressibility of
the electron liquid calculated for the uniform electron density in the
interstitial region outside muffin-tin spheres is generally a good
approximation to the compressibility of real metals. This general trend 
can be observed over a wide variety of metals covering
simple, transition and noble metals. The value of the inverse
compressibility $1/\kappa$ for transition metals is generally large owing to the
presence of d electrons, while the value of $1/\kappa$ for alkali metals is small.
It is very small particularly for sodium, potassium, and rubidium with the
lowest metallic densities.

We may therefore conclude that the electron liquid model is valid for
all metals
in the sense that the uniform electron density in the interstitial region
makes a predominant contribution to the inverse
compressibility $1/\kappa$ of metals. This implies that the metal cannot survive
if its uniform electron density in the interstitial region exceeds the
lowest possible critical density of the electron liquid where the
compressibility $\kappa$ becomes divergent and the system can no longer be
stable thermodynamically. This critical density corresponds to about
$r_s=5.25$. It has been generally observed that the uniform electron density
in the interstitial region of metals is somewhat enhanced in comparison
with the averaged density of valence electrons as a whole. 
Let us describe the uniform electron density in the interstitial region 
by the effective density parameter ${r_s}^*$. Then,
${r_s}^*<r_s$
for almost all metals. The values of
${r_s}^*$ for sodium, potassium, and rubidium are about
$3.66$, $4.36$, and $4.67$, respectively, while those of
$r_s$ are $3.96$, $4.87$, and $5.18$, respectively. For cesium with the lowest
observable
metallic density, the usual density parameter $r_s=5.57$ is beyond $5.25$, but its
effective density parameter ${r_s}^*$ is considered to be less than $5.25$.
  
  In fact, the LDA has borrowed the knowledge of the exchange-correlation
energy from the electron liquid beyond its applicability. One can
practically calculate its ground state energy beyond the critical density
resorting to a variety of approximation methods, but this does not implies
that the system stays stable. According to many-body theoretic techniques,
the exact theory should maintain the internal consistency between
self-energy and vertex parts and fulfill the Pauli principle. Since the
vertex part is closely related to the compressibility $\kappa$ of the system, this
internal consistency can no longer be maintained when it is divergent. In
this sense the electron liquid is no longer stable for $r_s>5.25$.
    
 In accordance with the situation above, we borrow the knowledge of the
effective interaction from the electron liquid for $r_s<5.25$. A practical
choice of the density parameter for the effective interaction probably will
be the parameter $r_s^*$ appropriate for the uniform electron density in the
interstitial region of metals calculated from the LDA. The best choice of
the density parameter is the optimization
that minimizes the calculated ground state energy in the present theory as
a function of the density parameter; $r_s^*$ will be a good
approximation to the optimized density parameter.
    
 The present correlation energy functional borrows the effective
interaction from the electron liquid. Therefore we have to treat separately the
functional derivative $\delta v_{eff}({\bf r'})/\delta n({\bf r})$ when we solve the OEP integral
equation in order to evaluate the exchange-correlation potential
$v_{xc}({\bf r})(=v_x({\bf r})+v_c({\bf r}) )$. An approximation appropriate for this purpose
will be that $\delta v_{eff}({\bf r'})/\delta n({\bf r})=\delta({\bf r-r'})dv_{eff}({\bf r})/dn$. 
The derivative $dv_{eff}({\bf r})/dn$ should be evaluated at the same density parameter as in the
effective interaction $v_{eff}({\bf r})$.
  
   So far we have discussed how to choose the density parameter for the
effective interaction $v_{eff}({\bf r})$ when it is intended for use in metals. 
Originally, the knowledge
of the correlation energy of the electron liquid should be applied to the
study of the electronic structure of metals. The LDA, if
applied to atoms, underestimates the exchange energy in magnitude by $14\sim 8\%$
and reversely overestimates the correlation energy by a factor of $2\sim2.5$ \cite{99,100}.
This is because the spurious self-interaction terms are not cancelled
and because the correlation energy of the
uniform electron liquid increases its magnitude like $\ln r_s$ Ry. per electron
at high densities.

In reality, the correlation energy of atoms is in magnitude comparable
with that of conduction electrons in metals, i.e., about $75 \sim 66$ mRy. per
electron. However, the ratio of the correlation energy to the exchange
energy is an order of magnitude larger in conduction electrons than in
atomic electrons; in atoms the accurate evaluation of the exchange energy
is indispensable before the correlation energy is allowed for.

The present theory gives the exact evaluation of the exchange energy
and moreover does not overestimate the correlation energy of atoms at all
since it borrows the knowledge of the effective interaction from the
electron liquid, not of the density dependence of the correlation energy. 
It should, however, be noted that the effective interaction $v_{eff}({\bf r})$ 
of the electron liquid, if applied to atoms, probably will lead to an underestimation at long
distances. This is because the screening in atoms is not perfect in
contrast to the electron liquid. If one defines the effective interaction
properly for atoms in a similar way as in the electron liquid, the ratio of
the Fourier transformed effective interaction $v_{eff}({\bf q})$ to the bare Coulomb
should tend to a finite value of order $1/\epsilon_0$, not to zero in the limit $q \to 0$
where $\epsilon_0$ is the static dielectric constant of atoms. Fortunately, the
small-wavenumber components of $v_{eff}({\bf q})$ are not expected to make an
important contribution to the correlation energy of atoms since the predominant part of the correlation
energy comes from virtual transitions to unoccupied localized orbitals with discrete spectra.
Similarly, the ratio of the effective
interaction properly defined for insulators and semiconductors should have
a finite value in the limit $q \to 0$ owing to the presence of the energy gap.

On the other hand, no qualitative difference in the intermediate- and
short-wavenumber components of the effective interaction may be expected
between the electron liquid and real atoms, insulators, and semiconductors.

To conclude, the best choice of the density parameter for the
effective interaction in the present theory is the optimization that
minimizes the calculated ground state energy as a function of the density
parameter, irrespective of the difference between of metals, insulators and
atoms. It is noteworthy that in the limit
of uniform density Kohn-Sham orbitals are reduced to plane waves and the
resulting ground state energy in the present theory takes its minimum when
the density parameter for the effective interaction amounts approximately
to $r_s=4.2$.

There is another method of choosing the density parameter for $v_{eff}({\bf r})$.
It utilizes the reference one-electron part of the virial theorem in terms
of DFT \cite{80,98}.
\begin{eqnarray}
2T_s+V_{es}+V_{ee}^x- \int d{\bf r}n({\bf r}){\bf r} \cdot {\rm grad}v_c({\bf r})=0,\label{eq:4.13}
\end{eqnarray}
where $V_{es} (V_{es}=V_{ne}+V_{ee}^H +V_{nn})$ is the total classical electrostatic energy
functional that consists of the nucleus-electron interaction energy $V_{ne}$,
the Hartree energy $V_{ee}^H$  and the
nucleus-nucleus interaction energy $V_{nn}$;
$V_{ee}^x$ stands for the exchange energy functional $(V_{ee}^x=E_x)$. The density
parameter for $v_{eff}({\bf r})$ can in principle be determined from the requirement
that Eq.(\ref{eq:4.13}) be satisfied. The minimization of the exact ground state energy
expression and the virial theorem should be fulfilled simultaneously at the exact
equilibrium inter-nucleus separation.

It should be noticed that the usual expression for the virial theorem, $2T+V=0$ can be
obtained from Eqs. (\ref{eq:4.12}) and (\ref{eq:4.13}); $T=T_s+T_c$ and $V$ stands for the total
potential energy of the system.

In the electron liquid with the neutralizing uniform positive
background correlation shifts the equilibrium density parameter from
$r_s=4.83$ to $4.2$. Correspondingly, the ground state energy is lowered from 
$-0.095$ to about $-0.154$ Ry. per electron. From the virial theorem with and without
correlation it follows that $2(T-T^{HF})+(V_{ee}-V_{ee}^{HF})=0$ where $T$ and $V_{ee}$ are the
exact kinetic and electron-electron interaction energies of the electron
liquid, respectively; the superscript HF is attached to the
corresponding Hartree-Fock values.

There is a second-order perturbation like relation \cite{54} between large wave number components of $T$ associated with the 
one-electron momentum distribution function $n({\bf k})$ and those of $V_{ee}$ associated with the static
structure factor $S({\bf k})$ due to short-range correlation, i.e.
$2T({\bf k})+V_{ee}({\bf k})=0$ for $|{\bf k}| \gg p_F$.

In real molecules and solids, on the other hand, the Hartree-Fock
model generally gives a fairly good approximation to the equilibrium
nucleus-nucleus separation. Correlation tends to increase the equilibrium
separation by only a small amount. The Hartree-Fock potential curve of diatomic molecules as
a function of the nucleus-nucleus separation, if shifted nearly uniformly
downward by an amount of the correlation energy, leads to the exact potential
curve in the neighborhood of the equilibrium separation. This implies that
$2(T-T^{HF})+(V_{ee}-V_{ee}^{HF})+(V_{ne}-V_{ne}^{HF})$ (or equivalently $-(V_{nn}-V_{nn}^{HF})$ ) nearly
vanishes or is a small positive value where $V_{ne}$ and $V_{nn}$ denote the
electron-nucleus and nucleus-nucleus interaction energies, respectively.

Besides the net correlational lowering in the electron-electron
interaction energy partially cancelled by the correlational increase in the
kinetic energy, the correlation-induced less screening of the nuclei
produces a further lowering in the electron-nucleus potential energy by
shrinking the electron distribution in the vicinity of the nuclei and
accordingly the lattice is somewhat expanded to reduce the nucleus-nucleus
interaction energy, relaxing the simultaneous increase in
the kinetic energy of the reference non-interacting system $T_s$ to some extent.

Thus, the correlational lowering in the electron-nucleus potential energy, which is
absent in the uniform electron liquid and hence cannot be properly
described in the LDA, plays a significant role in the accurate evaluation
of the equilibrium nucleus-nucleus separation as well as the electron
density $n({\bf r})$.

\section{RELATIONSHIP BETWEEN KOHN-SHAM EQUATIONS AND DYSON EQUATIONS}
In this section we first evaluate the quasiparticle energy dispersion of
the electron liquid with high accuracy. Next we give a detailed discussion about the relationship
between Kohn-Sham equations and Dyson equations in order to justify the
application of Kohn-Sham equations to the energy-band theory.
\subsection{Symmetric expression for the self-energy}
Sham and Kohn \cite{9} were the first who discussed the relationship between
Kohn-Sham equations and Dyson equations. In the system where the electron
density is slowly varying, they have found that in the neighborhood of the
Fermi level $\mu$ the quasiparticle energy dispersion $E({\bf p})-\mu$ can be related to
the Kohn-Sham energy dispersion $\epsilon^{KS}({\bf p})-\mu$ as
\begin{eqnarray}
E({\bf p})-\mu=\frac{\epsilon^{KS}({\bf p})-\mu}{ \int d{\bf r}\frac{m^*(n({\bf r}))}{m}|\varphi_{\bf p}({\bf r})|^2},\label{eq:5.1}
\end{eqnarray}
where $m^*(n({\bf r}))$ denotes the effective mass of the uniform electron liquid
appropriate for the local density $n({\bf r})$ and
$\varphi_{{\bf p}}({\bf r})$ is the Kohn-Sham orbital
specified by wavenumber ${\bf p}$.
    
 It is very difficult to give a qualitatively reliable evaluation of
the effective mass $m^*$ of the electron liquid over the entire region of
metallic densities. This is because the expansion of the self-energy with
respect to the screened Coulomb interaction converges very slowly in this
region. The effective mass ratio $m^*/m$ calculated to the first order in the
RPA screened interaction, giving the leading term in the high density
expansion of $m^*/m$ for small $r_s$ \cite{101}, takes the minimum around $r_s=0.7$, exceeds
unity around $r_s=2.3$ and thereafter slowly increases with $r_s$. The ratio
$m^*/m$ calculated to the second order stays less than unity until $r_s=6$, but
it has also the minimum around $r_s=1$ and thereafter increases slowly with
$r_s$. The very slow convergence of the effective mass ratio $m^*/m$ reflects
that correlation is much complicated in the region of metallic densities.
   
  In the language of many-body perturbation theory, this implies that
the determination of the quasiparticle energy dispersion requires the
accurate knowlegde of the self-energy $\Sigma_{\sigma}({\bf p},\epsilon)$ as a functional of the one-electron
Green's function $G_{\sigma '}({\bf p'},\epsilon')$ rather than a few lowest order expansion of $\Sigma_{\sigma}({\bf p},\epsilon)$ in
the screened Coulomb interaction, though screening is of course necessary
to overcome the exchange difficulty.  As is proved in a celebrated book by
Noziers, any variation in the self-energy $\Sigma_{\sigma}({\bf p},\epsilon)$ occurs only through a
variation in each
constituent $G_{\sigma '}({\bf p'},\epsilon')$ of the self-energy $\Sigma_{\sigma}({\bf p},\epsilon)$. This is the reason why
the accurate knowledge of $\Sigma_{\sigma}({\bf p},\epsilon)$ as a functional of $G_{\sigma '}({\bf p'},\epsilon')$ is
essential for the evaluation of the quasi-particle energy dispersion.
   
  Yasuhara and Takada \cite{24} have introduced the very expression for the
self-energy $\Sigma_{\sigma}({\bf p},\epsilon)$ that is symmetric with respect to each consitituent $G$
in it.
\begin{eqnarray}
&&\Sigma_{\sigma}({\bf p},\epsilon; [G])=\sum_{\sigma '}\int\frac{d{\bf p'}}{(2\pi)^3}\int d \epsilon ' \sum_{n=1}^{\infty}\frac{1}{2n-1}\nonumber\\
&&~~~~~~~~~\times I^{(n)\sigma \sigma '}({\bf p},\epsilon;{\bf p}',\epsilon'; [G])G_{\sigma '}({\bf p'}, \epsilon '),\nonumber\\
&&I^{\sigma \sigma '}({\bf p},\epsilon;{\bf p}',\epsilon '; [G])=\sum_{n=1}^{\infty}I^{(n)\sigma \sigma '} ({\bf p},\epsilon;{\bf p}',\epsilon '; [G]),\nonumber\\
\label{eq:5.2}
\end{eqnarray}
where $I^{(n)\sigma \sigma'}({\bf p},\epsilon;{\bf p}',\epsilon'; [G])$ denotes the contribution to the particle-hole
irreducible interaction from all the possible $n$-th order skeleton diagrams.
According to this symmetric expression, one can easily check that the
functional derivative
$\delta \Sigma_{\sigma}({\bf p},\epsilon; [G])/\delta G_{\sigma'}({\bf p}',\epsilon')$
is equivalent to the particle-hole irreducible
interaction
$I^{\sigma \sigma'}({\bf p},\epsilon;{\bf p}',\epsilon'; [G])$, as is proved in standard textbooks on many-body
theory. The interaction $I^{\sigma \sigma'}({\bf p},\epsilon;{\bf p}',\epsilon'; [G])$
is the value of the full particle-hole irreducible interaction 
$I^{\sigma \sigma'}({\bf p},\epsilon; {\bf p}',\epsilon'; {\bf q},\omega; [G])$
in the limit of ${\bf q}=\omega=0$. The symmetric expression has the striking merit of
visualizing that a great deal of cancellation occurs systematically between
higher-order contributions due to the Pauli principle. This is
indispensable to the qualitative and quantitative evaluation of the
quasiparticle energy dispersion as well as the effective mass of the
electron liquid in the strongly correlated density region.

A great deal of cancellation can also be observed between the
wavenumber dependence and the energy dependence of the self-energy
$\Sigma({\bf p},\epsilon; [G])$ in the determination of the quasiparticle energy dispersion $E({\bf p})$.
This can be traced back to the dynamical nature of screening in the electorn liquid.
Both of the wavenumber dependence and the energy dependence are
comparatively large in magnitude, but they cancel each other for the most
part in the determination of $E({\bf p})$ and $m^*$.

     Since the early stage of the study of the electron liquid \cite{102,103}, it has been
recognized that the first iterative solution for the quasiparticle energy
dispersion, i.e., $E({\bf p})=\epsilon({\bf p})+\Sigma_{\sigma}({\bf p},\epsilon({\bf p}); [G^0])$ 
is a good approximation to the
solution of the fully self-consistent equation, $E({\bf p})=\epsilon({\bf p})+\Sigma_{\sigma}({\bf p},E({\bf p}); [G])$.
This is because there is a great deal of cancellation between the
higher-order iterative correction and the correction made by replacing
$\Sigma_{\sigma}({\bf p},\epsilon; [G^0])$ with $\Sigma_{\sigma}({\bf p},\epsilon; [G])$; 
the partially self-consistent solution with $\Sigma_{\sigma}({\bf p},\epsilon; [G^0])$, the solution of 
$\epsilon=\epsilon({\bf p})+\Sigma_{\sigma}({\bf p},\epsilon; [G^0])$ becomes worse. Almost no
difference can be expected between the first iterative and the 
fully self-consistent solutions around the Fermi level since the electron
liquid is a many-body system which consists of electrons with the same character.

     The self-energy as a functional of $G^0$, $\Sigma_{\sigma}({\bf p},\epsilon; [G^0])$ can be written as \cite{43}
\begin{eqnarray}
\Sigma_{\sigma}({\bf p},\epsilon; [G^0])=\frac{ (e^2)^{ \frac{1}{2} } }{2}\int de^2(e^2)^{-\frac{3}{2}}~~~~~~~~~~~~~~~\nonumber\\
\times\sum_{\sigma'}\int\frac{d{\bf p}'}{(2\pi)^3}\int d \epsilon '
I^{\sigma \sigma'}({\bf p},\epsilon;{\bf p}',\epsilon'; [G^0])G^0_{\sigma'}({\bf p}',\epsilon').\nonumber\\ \label{eq:5.3}
\end{eqnarray}
where the sum over $n$ with the factor $1/(2n-1)$ in Eq. (\ref{eq:5.2}) is transformed
into the coupling constant integral since $G^0$ does not depend on the
coupling constant.
All diagrams for the particle-hole irreducible interaction $I^{\sigma \sigma'}({\bf p},\epsilon;{\bf p}',\epsilon'; [G^0])$ 
can be classified into three different types, improper, doubly irreducible, 
and proper and reducible \cite{24} (see Fig. 2). From this classification
it is evident that there is a great deal of systematic cancellation among
higher order contributions of $I^{\sigma \sigma}({\bf p},\epsilon;{\bf p}',\epsilon'; [G^0])$ 
due to the Pauli principle. In the practical calculations we have employed an approximate
form of $I^{\sigma \sigma}({\bf p},\epsilon;{\bf p}',\epsilon'; [G^0])$, 
in which the above cancellation due to the
Pauli principle is allowed for systematically up to higher orders; on the
other hand, we have put the contribution from $I^{\sigma-\sigma}({\bf p},\epsilon;{\bf p}',\epsilon'; [G^0])$ as a
constant since it is rather small; we have chosen the constant by fitting
the value of $\Sigma({\bf p},\epsilon({\bf p}); [G^0])$ at $p=p_F$ to the exchange-correlation contribution
of the chemical potential $\mu_{xc}$.
\begin{figure}[t]
  \includegraphics[width=7.5cm,keepaspectratio]{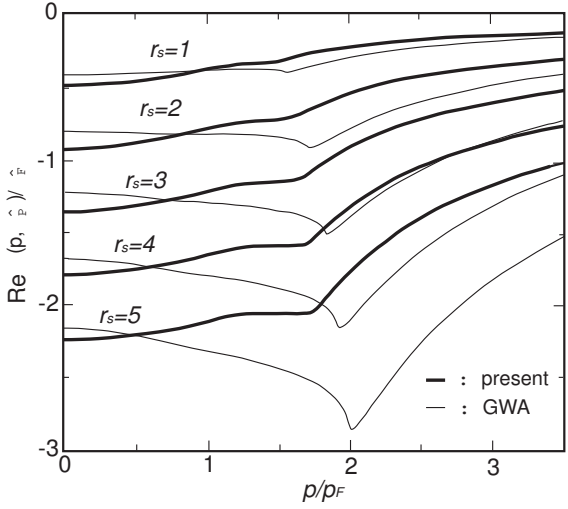}
\caption{\label{fig11} The real part of the self-energy correction calculated from the
present theory, ${\rm Re}\Sigma({\bf p},\epsilon ({\bf p}); [G^0])$ in units of the Fermi energy $\epsilon_F$ as a
function of $q/p_F$ for various values of $r_s$. For comparison the correction
calculated from the GW approximation is also drawn.}
\end{figure}

     Fig. 11 shows the real part of the self-energy correction calculated
from the present theory in units of the Fermi energy $\epsilon_F$ for various values of $r_s$. 
For comparison the correction calculated from the GW approximation
is also drawn in the figure. A clear difference in the wavenumber
dependence of ${\rm Re}\Sigma_{\sigma}({\bf p},\epsilon({\bf p}); [G^0])$ can be seen between $0< p< p_F+q_c$ and $p_F+q_c< p$
where $q_c$ is the cutoff wavenumber of plasmon excitations. The present
self-energy correction ${\rm Re}\Sigma_{\sigma}({\bf p},\epsilon({\bf p}); [G^0])$ maintains a small positive slope
over a range of $0< p< p_F+q_c$ throughout the entire region of metallic
densities. The slope of ${\rm Re}\Sigma_{\sigma}({\bf p},\epsilon({\bf p}); [G^0])$ at the Fermi wavenumber $p_F$ is
closely related to the effective mass $m^*$ and is a slowly increasing
function of $r_s$. The value of the mass ratio $m^*/m$ estimated from the slope
is consistent with the result in a previous study of the effective mass \cite{16,40,41}.
The mass ratio $m^*/m$ is a slowly decreasing function of $r_s$ and is shifted
from unity about $7\sim8\%$ even at the lowest metallic densities. This means
that the wavenumber dependence of $\Sigma_{\sigma}({\bf p},\epsilon; [G])$ somewhat overcomes in
magnitude the energy dependence of $\Sigma_{\sigma}({\bf p},\epsilon; [G])$ in the determination of the
quasiparticle energy $E({\bf p})$ in the neighborhood of the Fermi level.
  
  For $p>p_F+q_c$ , on the other hand, the self-energy correction
${\rm Re}\Sigma_{\sigma}({\bf p},\epsilon({\bf p}); [G^0])$
behaves approximately like $-1/p$ in a quite different way from for
$0< p< p_F+q_c$. The critical wavenumber near $p_F+q_c$ corresponds approximately to
the energy $\mu+\hbar\omega_{pl}$ where $\hbar\omega_{pl}$ is the plasmon excitation energy. The
remarkable difference in the wavenumber dependence of the self-energy
correction can be ascribed to the fact that screening does work only for
$0<p<p_F+q_c$ in the electron liquid.

     The GW approximation gives a fairly good evaluation of the magnitude
of the self-energy correction, but it fails to give a qualitatively correct
description of its derivative for $0<p<p_F+q_c$ over almost the entire region
of metallic densities. The appearence of a dip around $p_F+q_c$ in both cases
is ascribed to the use of ${\rm Re}\Sigma_{\sigma}({\bf p},\epsilon; [G^0])$ in place of 
${\rm Re}\Sigma_{\sigma}({\bf p},E({\bf p}); [G])$. The dip should be destroyed under the influence of a coupling between plasmon
excitations and multipair excitations.

     From these observations we may conclude that the self-energy
correction to the quasiparticle energy dispersion $E({\bf p})$ is fairly well
approximated by its value at the Fermi wavenumber $p_F$ for a range of
wavenumbers $0<p<p_F+q_c$ throughout the entire region of metallic densities.
In the limit of uniform density the Kohn-Sham energy dispersion becomes
$\epsilon^{KS}({\bf p})=\epsilon({\bf p})+\mu_{xc}$. The ratio 
$\{{\rm Re}\Sigma({\bf p},\epsilon({\bf p}); [G^0]-\mu_{xc}\}/\mu_{xc}$ is less than about 10$\%$
for $0<p<p_F+q_c$. Therefore we may say that Kohn-Sham energies are a rather
good approximation to quasiparticle energies so far as excitation energies
relative to the Fermi level are less than about $\hbar\omega_{pl}$.

The conclusion above will probably apply to real metals. This is
because the large cancellation between the wavenumber dependence and the
energy dependence of the self-energy $\Sigma({\bf p},\epsilon; [G])$ in the determination of the
quasiparticle energy dispersion $E({\bf p})$ below the critical energy near $\mu+\hbar\omega_{pl}$
can be traced back to the dynamical nature of screening that is associated
with the presence of the Fermi surface. 

An energy-band calculation of
sodium \cite{104} performed in the Yasuhara-Takada approximation \cite{24} much more advanced
than the usual GW approximation has actually revealed that the occupied
band width of sodium is increased by an amount less than 10$\%$, compared with
the conventional LDA result.

Very recently, R. Maezono, M.D. Towler, Y. Lee, and R. J. Needs \cite{105} have
performed the diffusion Monte Carlo (DMC) energy-band calculation of sodium
to find confidently that its occupied band width is increased to the almost
same extent as the present result. It seems that the exchange difficulty of
Hartree-Fock equations associated with metals as well as the
electron liquid has long been believed
to be resolved completely by the RPA, but in fact it is not the case, i.e.,
the RPA only removes the infinite derivative at $p_F$.
It is concluded that the present many-body theoretic
study and the above DMC calculation have cooperated to resolve in a truly
complete sense the exchange difficulty of the Hartree-Fock theory at
realistic metallic densities.
\subsection{Relation between $\Sigma({\bf r,r'},\epsilon)$ and $v_{xc}({\bf r})$}
Dyson equations and Kohn-Sham equations are derived from two different
variational principles concerning the same ground state energy $E_g$ ,
$\delta E_g/\delta G({\bf r,r'},\epsilon)=0$ and $\delta E_g/dn({\bf r})=0$, respectively, where $E_g$ can be considered
to be a functional of the one-electron Green's functional $G({\bf r,r'},\epsilon)$ or
another functional of the electron density $n({\bf r})$. The two different types of
equations can in principle give the same exact electron density $n({\bf r})$,
chemical potential and ground state energy of a many-electron system.

Sham and Schluter \cite{106,107}  have derived an exact expression for the
exchange-correlation energy functional $E_{xc}$ by means of many-body theoretic 
technique based on the Green's function formalism and related the exchange-correlation 
potential $v_{xc}({\bf r})$ in DFT to the self-energy $\Sigma({\bf r,r'},\epsilon; [G])$. This
Sham-Schluter equation implies that the electron density $n^{KS}({\bf r})$ evaluated
from the Kohn-Sham Green's function $G^{KS}({\bf r,r'},\epsilon )$ as the unperturbated Green's
function in the many-body perturbation expansion is identical to the
electron density $n({\bf r})$ calculated from the exact one-electron Green's
function $G({\bf r,r'},\epsilon )$.
    
 Grabo, Kreibich, Kurth and Gross \cite{8} have discussed the OEP method from
the viewpoint of many-body perturbation expansion. They have then pointed
out that the higher order contribution of the exchange-correlation energy
functional beyond the Hartree-Fock approximation, $E^{(n)}_{xc} (n \geqq 2)$ depends not
only on occupied and unoccupied Kohn-Sham orbitals and Kohn-Sham energies
but also on the exchange-correlation potential $v_{xc}({\bf r})$. This implies that
the resulting OEP integral equation involves the functional derivative
$\delta v_{xc}({\bf r})/\delta n({\bf r'})$ as well as the potential $v_{xc}({\bf r})$ and is much more difficult
to solve as a result.

     Casida \cite{108} has started with an approximate expression for the ground state
energy functional. In the expansion of the self-energy functional using
skeleton diagrams he has omitted those terms which arise from the coupling
constant integral of the constituent renormalized Green's functions,
or equivalently those terms which vanish if the exact one-electron
Green's function $G$ is replaced by the unperturbed Kohn-Sham Green's
function $G^{KS}$ in the exact ground state energy functional for the
variational principle. Using this approximate ground state energy
functional $E_0[G^{KS}]+\Phi[G^{KS}]$, Casida has generalized the exchange-only OEP method by Sharp and
Horton \cite{109,110} to include correlation. The resulting OEP integral equation has
turned out to be identical to the first iterative form of the Sham-Schluter
equation (see Eq.(\ref{eq:5.4})).
Casida has thus shown that the exchange-correlation potential $v_{xc}({\bf r})$ in DFT
can be interpreted to be the variationally best optimized local
approximation to the exchange-correlation part of the self-energy $\Sigma({\bf r,r'},\epsilon; [G])$.

     We would like to stress here that Casida's expression for the ground
state energy functional $E_0[G^{KS}]+\Phi[G^{KS}]$ is a sufficiently accurate approximation for the
practical purpose of studying the electronic structure of solids. This may
be understood from the fact that the same approximation, if applied to the
uniform electron liquid, gives an excellent description of long-,
intermediate-, and short-range correlations and furthermore obeys the Pauli
principle; it can also give a highly accurate evaluation of the ground
state energy for the electron liquid. This is because the self-energy
functional employed is exact.

     We may say that Kohn-Sham equations assume the complete cancellation
between the nonlocality (or the wavenumber dependence) and the energy
dependence of $\Sigma({\bf r,r'},\epsilon; [G])$ in the evaluation of quasiparticle energies.
This assumption is rather good, as we have already discussed in Sec. V A.

    As for the energy gap of semiconductors, the gap of Kohn-Sham equations
is related to that of Dyson equations as $E_{gap}^{KS}+\Delta_{xc}=E_{gap}$ where $\Delta_{xc}$ denotes the
discontinuity of the exchange-correlation potential $v_{xc}({\bf r})$ \cite{111,112}. According to an
accurate
Hartree-Fock calculation in DFT \cite{86}, the relation $E_{gap}^{HFKS}+\Delta_{x}=E_{gap}^{HF}$ has been
ascertained for silicon within an accuracy of $3\%$. It has been reported that
the value of $E_{gap}^{HFKS}$ is a good approximation of the experimental gap and the
discontinuity $\Delta_x$ amounts to about twice the experimental gap. This implies
that $\Delta_x$ must be for the most part cancelled by $\Delta_c$. According to standard
many-body perturbation theory, the well-known overestimation in the gap of
semiconductors calculated from usual Hartree-Fock equations can be remedied
by the screened exchange potential. In the framework of DFT, on the other
hand, the same screening effect can be represented by a great deal of
cancellation between $\Delta_x$  and $\Delta_c$. Probably, the total 
discontinuity $\Delta_{xc}(=\Delta_x+\Delta_c)$ 
will be small but positive by analogy with the difference in the energy
dispersion of the electron liquid between Dyson equations and Kohn-Sham
equations. Kohn-Sham equations overestimate the density of states at the
Fermi level by less than $10\%$ through the scaling factor $m/m^*$.
\subsection{Correspondence between two self-consistencies}
In an attempt to interpret the extraordinarily enhanced effective mass of
heavy fermion systems using the Hubbard model, a number of authors \cite{113} have
exaggerated the role of the energy dependence of the self-energy in the
determination of the quasiparticle energy dispersion around the Fermi
level, neglecting its wavenumber dependence. In fact, there is
a great deal of cancellation between the wavenumber dependence and the
energy dependence of the self-energy. The former somewhat overcomes the
latter in magnitude as we have elucidated in Sec. V A. This characteristic
of the self-energy can be traced back to the dynamical nature of screening
and hence is expected to apply to all metals as far as the Fermi
surface is well defined. Then we have to ascribe the origin of the
extraordinarily enhanced effective mass to a different charateristic of the
self-energy in Dyson equations.

     In this subsection we would like to point out a characteristic of the
self-energy  peculiar to nonuniform real metals as a promising candidate
for the origin of the
extraordinarily enhanced effective mass. In transition metals and rare
earth metallic
compounds the Fermi surface is constructed from a few kinds of electrons
with different orbital characters, in contrast with the electron liquid
which consists of a single kind of electrons with the same character.

     Consider the self-energy $\Sigma_{\sigma}({\bf r,r'},\epsilon; [G])$
 entering Dyson equations. It is
a nonlocal and energy-dependent operator, but it is, at the same time, a
functional of the self-consistent Green's function $G$. In the uniform
electron liquid there is no essential difference in the neighborhood of the
Fermi level between the first iterative and the fully self-consistent
solutions for the quasiparticle energy dispersion. In the case of rare
earth metallic compounds, on the other hand, there is a very strong
possibility that being a functional of the self-consistent Green's function
$G$ may cause a drastic change in the quasiparticle energy dispersion
particularly in the immediate vicinity of the Fermi level due to the
presence of f electrons and gives an extraordinarily sharp enhancement in
the density of states. Probably, such a drastic change in the dispersion
will be caused by the correlation-induced reconstruction of quasiparticle
states in the immediate vicinity of the Fermi level and of course involve a
significant change in the electron density $n({\bf r})$ of the system.

     According to Casida, the exchange-correlation potential $v_{xc}({\bf r})$ in
Kohn-Sham equations can be interpreted to be the variationally best optimized local
approximation to the self-energy $\Sigma({\bf r,r'},\epsilon; [G])$. Kohn-Sham equations assume
the complete cancellation between the wavenumber dependence (non-locality) and
the energy dependence of the self-energy, but still they maintain its
essential property as a functional of the self-consistent $G$ in the form of
the exchange-correlation potential $v_{xc}({\bf r})$ as a functional of the
self-consistent electron density $n({\bf r})$. In other words, the implicit
dependence of $\Sigma({\bf r,r'},\epsilon; [G])$ on the self-consistent $G$ is
in one-to-one correspondence to the implicit dependence of
$v_{xc}({\bf r}; [n])$ on the self-consistent $n({\bf r})$ in the sense that the latter
occurs only from the former. This may be seen from the first iterative form of the 
Sham-Schluter equation which is identical to the OEP integral equation.
\begin{eqnarray}
&&\int d{\bf y}d{\bf y'}\int\frac{d\omega}{2\pi}G^{KS}({\bf r,y},\omega)\delta({\bf y-y'})\nonumber\\
&&~~~~~~~~~~~~~~~~~~~~~~~~~\times v_{xc}({\bf y}; [n])G^{KS}({\bf y',r},\omega)\nonumber\\
&=&\int d{\bf y}d{\bf y'}\int\frac{d\omega}{2\pi}G^{KS}({\bf r,y},\omega)\Sigma({\bf y,y'},\omega; [G^{KS}])\nonumber\\
&&~~~~~~~~~~~~~~~~~~~~~~~~~~~~\times G^{KS}({\bf y',r},\omega),\label{eq:5.4}
\end{eqnarray}
where $G^{KS}({\bf r,y},\omega)$ denotes the unperturbed Kohn-Sham Green's function in the
many-body perturbation expansion.

     We may therefore expect that such a correlation effect on
quasiparticle energies as involves a change in the electron density $n({\bf r})$ of
the system probably will be for the most part taken into account in the
framework of Kohn-Sham equations in DFT. This is parallel to the
expectation that there will probably be no significant difference between
the Kohn-Sham Fermi surface and the true Fermi surface, though strictly
speaking the two Fermi surfaces are not identical \cite{114,115}. The
one-to-one correspondence between the two self-consistencies, i.e.,
$\Sigma({\bf r,r'},\epsilon; [G^{KS}])$ as a functional of the self-consistent $G^{KS}$ and $v_{xc}({\bf r};[n])$
as a functional of the self-consistent $n({\bf r})$ will persist even for strongly
anisotropic systems such as heavy fermion systems.

     In Sec. IV we have predicted that the correlation-induced
reconstruction of Kohn-Sham orbitals in the vicinity of the Fermi level or
the energy gap will be
essential to the theoretical interpretation of the electronic structure of
the so-called strongly correlated electron systems. This prediction is
based on the one-to-one
correspondence between the two self-consistencies we have just mentioned
above.

     To conclude, if the correlation energy functional to be used is
sufficiently accurate, Kohn-Sham equations are expected to give a
good approximation to quasiparticle energies as far as they do not exceed a
critical energy near $\mu+\hbar\omega_{pl}$. The orbital-dependent correlation 
energy functional we have
proposed in Sec. IV B to be used in combination with the orbital-dependent
exchange energy functional probably will be sufficiently accurate to promote development
in the band theory of solids.
\section{CONCLUDING REMARKS}
One of the most important subjects in solid state physics at present is
exploitation of  more accurate methods for treating correlation in the
theoretical study of the electronic structure of solids. For this purpose
we have constructed an orbital-dependent correlation energy functional
$E_c[\{\varphi_i\}]$ to be used in combination with the orbital-dependent exchange
energy functional $E_x[\{\varphi_i\}]$ in density-functional theory (DFT) with the help
of detailed knowledge of long-, intermediate-, and short-range correlations
of the electron liquid, not its density dependence of the correlation
energy. This is a sophisticated but feasible method which may develop the
band theory to a new advanced stage beyond the LDA. The present correlation
energy functional $E_c[\{\varphi_i\}]$ in its construction is valid for tightly binding
electrons as well as for nearly free electrons.

     Generally, effects of correlation on the electronic structure of
solids are so  complicated and delicate that in order to obtain a reliable
conclusion one has to take account of various aspects of correlation
systematically starting from the original Coulomb interaction between
valence electrons participating in cohesion of solids instead of resorting
to intuitive simplifications such as the Hubbard model.

     Two aspects of long-range correlation arising from long-range parts of
the Coulomb interaction in the electron liquid, i.e., the existence of
plasmons as a collective excitation and screening of the Coulomb
interaction between individual excitations (quasiparticles) as its
complementary effect in the lower energy region apply to real metals though
both are somewhat modified by band structure effects. The RPA is the simplest
approximation to describe these aspects and its screening is
indispensable to overcome the exchange difficulty. However, it is important
to recognize that the qualitative and quantitative evaluation of the
quasiparticle energy dispersion $E({\bf p})$ and the effective mass $m^*$ in the
region of metallic densities requires a systematic inclusion of
higher-order perturbation terms obeying the Pauli principle in the expansion
of the self-energy operator in the screened Coulomb interaction. This is
because long-, intermediate-, and short-range correlations occurring in this
region are very much complicated under the constraint of the Pauli
principle, in marked contrast to the situation in the high density region
where long-range correlation in the RPA is predominant. In the evaluation
of $E({\bf p})$ and $m^*$ of  the electron liquid, a great deal of cancellation is
observed between the wavenumber dependence and the energy dependence of the
self-energy operator. This can be interpreted as the dynamical nature of
screening and hence we may expect that this cancellation will be
observed in real metals so far as the Fermi surface is well defined.

     No qualitative difference may be expected in the neighborhood of the
Fermi level or the energy gap between the quasiparticle energy dispersion
calculated from Dyson equations and the one-electron energy dispersion from
Kohn-Sham
equations. The quantitative difference is less than $10\%$ so far as we know.
Then we may conclude that the use of Kohn-Sham equations will be valid for
further development of the band theory.

     The reason for the unexpected success of the LDA may be ascribed to
the long-range nature of the Coulomb interaction between electrons, which
tends to cancel the difference between a real metal and its locally uniform
electron liquid model in the quantitative evaluation of various
electron-electron interaction effects.

     Except for the correlated properties coming from the electron liquid,
what new correlation is the most important in real metals at all?  We
answer this question from the viewpoint of DFT. The present $E_c[\{\varphi_i\}]$ is
valid for both nearly-free and  tightly-binding electrons since it assumes
an orbital-dependent second-order perturbation like form in which all
effects of higher-order perturbation terms are represented by an effective
interaction defined for the electron liquid in the region of metallic
densities. A striking merit of the present $E_c[\{\varphi_i\}]$ is that it makes the meaning
of correlation clear against the Hartree-Fock approximation in the
framework of DFT and that it suggests how the electron density $n({\bf r})$ of the
system may be redistributed under the influence of correlation. The present
$E_c[\{\varphi_i\}]$ hybridizes Hartree-Fock Kohn-Sham orbitals in order to produce
self-consistently reconstructed Kohn-Sham orbitals through the correlation
potential $v_c({\bf r})$. The important point is that the present $v_c({\bf r})$ probably
gives rise to less screening of the attractive nuclear potential at short
distances to enhance the electron density $n({\bf r})$ in the immediate vicinity of
individual nuclei in solids.

    The electronic structure of the so-called strongly correlated electron
systems such as heavy fermions and Mott insulators might be interpreted in
terms of self-consistently reconstructed Kohn-Sham orbitals in the
neighborhood of the Fermi level or the energy gap.

     The expression "strongly correlated" has been extensively used in the
sense that the electronic structure is beyond the scope of the LDA. In
fact, the magnitude of the spin-antiparallel pair correlation function
$g^{\uparrow \downarrow}({\bf r,r'})$ for conduction electrons in cesium 
will be most reduced at
short separations. Correctly, we should state that the LDA fails to give
the proper description of correlation-induced reconstruction of Kohn-Sham
orbitals. This reconstruction may be interpreted as a natural consequence
of the fact that short-range correlation between valence electrons
participating in cohesion of solids induces inhomogeneity in the electron
density $n({\bf r})$ associated with the presence of the nuclei, or equivalently
short-range correlation between valence electrons significantly reduces the
Hartree- or Hartree-Fock
screening of the attractive nuclear potential at short distances to enhance
the electron density $n({\bf r})$ at the positions of individual nuclei in solids.
This correlation-induced reconstruction of Kohn-Sham orbitals involving the
enhancement of $n({\bf r})$ at the positions of individual nuclei is precisely the
most important correlation that the LDA fails to describe.

      The present theoretical treatment of correlation will manifest the
true value of DFT in the study of the electronic structure of solids. 
The present theory requires highly complicated
self-consistency because of the orbital-dependence of the
exchange-correlation energy functional $E_{xc} (=E_x+E_c)$. Much technical
difficulty is expected in its practical application. The most difficult
problem will be the accuracy required in solving the OEP integral equation
to evaluate the exchange-correlation potential $v_{xc}({\bf r})(=v_x({\bf r})+v_c({\bf r}))$. 
The KLI method of approximately solving the OEP integral equation may be
hopefully applied to the present orbital-dependent correlation energy
functional.
     It is not so difficult to construct the spin-polarized version of the
present theory.
\begin{acknowledgements}
One of the authors (H.Y.) would like to thank Professor Walter Kohn for
stimulating discussions about several fundamental problems of electron
correlation during his stay in Sendai in July of 2002. The authors wish to
thank President Akira Hasegawa in Niigata University for his critical
reading of the manuscript. One of the authors (H.Y.) is grateful to
Professor Komajiro Niizeki for his steady encouragement during the
preparation of the manuscript as well as for many helpful discussions about
electron correlation. He also would like to acknowledge Professor Kouichi
Ohno for his helpful discussion about the Hartree-Fock screening of the
nuclei in molecules. One of the authors (M.H.) is grateful to Professor
Tadao Kasuya for his instructive comments on heavy fermion systems. He also
would like to acknowledge Dr.Takahiro Maehira and Professor Katsuhiko
Higuchi for their helpful comments.
\end{acknowledgements}

\end{document}